\documentclass[12pt]{iopart}

\usepackage{bm}
\usepackage{graphicx}
\usepackage{caption}
\usepackage{subcaption}

\usepackage{cite}

\newcommand{\MATLAB}{\textsc{Matlab}\xspace}
\usepackage{xspace}

\usepackage{xcolor}

\usepackage{algorithm}
\usepackage{algorithmic}

\usepackage{iopams}
\usepackage{hyperref}
\usepackage[capitalize,noabbrev]{cleveref}
\begin{document}

\title[Damage identification in FML using Bayesian Analysis with MOR]{Damage Identification in Fiber Metal Laminates using Bayesian Analysis with Model Order Reduction}

\author{Nanda Kishore Bellam Muralidhar$^{1}$, Carmen Gräßle$^{2}$, Natalie Rauter$^{3}$, Andrey Mikhaylenko$^{3}$, Rolf Lammering$^{3}$, Dirk A. Lorenz$^{1}$}

\address{$^{1}$ Institute for Analysis und Algebra, Technische Universität Braunschweig, Universitätsplatz 2, 38106 Braunschweig, Germany}
\address{$^{2}$ Institute for Partial Differential Equations, Technische Universität Braunschweig, Universitätsplatz 2, 38106 Braunschweig, Germany}
\address{$^{3}$ Institute for Mechanics, Helmut-Schmidt-University/University of the Federal Armed Forces Hamburg, Holstenhofweg 85, 22043 Hamburg, Germany;}
\ead{n.bellam-muralidhar@tu-braunschweig.de}
\vspace{10pt}

\begin{abstract}

Fiber metal laminates (FML) are composite structures consisting of metals and fiber reinforced plastics (FRP) which have experienced an increasing interest as the choice of materials in aerospace and automobile industries. Due to a sophisticated built up of the material, not only the design and production of such structures is challenging but also its damage detection. This research work focuses on damage identification in FML with guided ultrasonic waves (GUW) through an inverse approach based on the Bayesian paradigm. As the Bayesian inference approach involves multiple queries of the underlying system, a parameterized reduced-order model (ROM) is used to closely approximate the solution with considerably less computational cost. The signals measured by the embedded sensors and the ROM forecasts are employed for the localization and characterization of damage in FML. In this paper, a Markov Chain Monte-Carlo (MCMC) based Metropolis-Hastings (MH) algorithm and an Ensemble Kalman filtering (EnKF) technique are deployed to identify the damage. Numerical tests illustrate the approaches and the results are compared in regard to accuracy and efficiency. It is found that both methods are successful in multivariate characterization of the damage with a high accuracy and were also able to quantify their associated uncertainties. The EnKF distinguishes itself with the MCMC-MH algorithm in the matter of computational efficiency. In this application of identifying the damage, the EnKF is approximately thrice faster than the MCMC-MH. 

\end{abstract}

%
\vspace{2pc}

\textbf{Main contributions:} This work focuses on the identification of damages in fiber metal laminates, where the damage is characterized by three parameters. The underlying high-fidelity forward model as well as the global parametric reduced-order model are taken from \cite{Bellam-Muralidhar-NK} and shortly recalled. The numerical experiments in this paper use the global parametric reduced-order model with the setting specified as in \cite[Section 5]{Bellam-Muralidhar-NK}. For the estimation of the damage parameters, we apply the existing methods Markov Chain Monte-Carlo (MCMC) based Metropolis-Hastings algorithm (see e.g.\ \cite{Metropolis, Hastings}) and Ensemble Kalman Filter technique (EnKF) (see e.g.\ \cite{Evensen}). Thus, the main contributions lie in bringing together and applying the well-known inference approaches with the global parametrized reduced-order model from \cite{Bellam-Muralidhar-NK} to the specific application of damage parameter estimation in fiber metal laminates using guided ultrasonic waves and analyzing numerically the accuracy and efficiency of the approaches for different noise levels as well as providing numerical estimates for the uncertainty.

\section{Introduction}
\label{sec:intro}
Reducing the weight of a product is often one of the central objectives in applications especially in aerospace structures. Fiber reinforced plastics (FRPs), owing to their high strength and stiffness to weight ratio, have been the first choice of material to satisfy this objective. However, the load bearing behavior, damage tolerance or impact-related properties of FRPs are not always desirable. These issues are addressed to a greater extent by fiber metal laminates (FMLs), first developed in 1980s, which partially combine the benefits of both metals and composites. The complicated structure of FMLs has made their design, fabrication as well as the repair and maintenance an intricate process. High performance structural FMLs are susceptible to inherent defects, such as matrix cracking and delamination. Such defects could either occur during their manufacturing process or in operation. 

Non-destructive evaluation (NDE) techniques are usually used to detect  damages in composites. Nonetheless, conventional methods like ultrasonic testing, computed tomography, laser shearography and X-ray radiography, are relatively complex, time and cost-intensive. Structural health monitoring (SHM) is an automated continuous on-board monitoring of a structure's integrity during operation using the sensors embedded within the structure \cite{Farrar}. Numerous SHM techniques have been developed since its inception. Guided ultrasonic wave (GUW) based SHM in FML is one of those methods with a great potential to identify the damage \cite{David-West}. The change in propagation behavior of GUW upon its interaction with a damage could be recorded by the integrated sensors \cite{Lammering-R, Su-Z, Guy-P}. The altered signals are then processed by efficient numerical methods to identify the damage. The damage identification procedure can be classified into: (a) model-based damage analysis and (b) model-free damage analysis \cite{Farrar}. The former approach \cite{Douglass-AC, Reed-H, Ng-CT, He-S} involves a model that mimics the underlying system, the propagation of GUW in an FML with a damage, along with the measurement data while the latter is purely driven on the measured data by the embedded sensors \cite{Cha-YJ, Atha-DJ, Dung, Wei-F, Abdeljaber-O}. 

This research work focuses on a finite element model (FEM) based analysis that utilizes the sensor measurement data for the damage identification. As the forward simulations are carried out at high frequencies, a fine spatial and temporal discretization of the model is required. This consequently results in a large number of degrees of freedom to be solved and thereby, a huge computational cost. Considering the uncertainties due to several endo-/exogenous factors, e.g.\ sensor measurement noise, model disparities and environmental variability, in this paper, the stochastic Bayesian inference approach is employed for the damage characterization \cite{Calvetti}. This procedure demands multiple solves of the expensive forward model which is often practically intractable for most of applications. Therefore, a reduced-order model (ROM), approximate to the high-fidelity (HiFi) model of the system, is used in the inference process for damage identification. Since parameter variations are inevitable in such inference methods, the authors have developed and validated a parameterized ROM \cite{Bellam-Muralidhar-NK} for the system and settings considered in this paper. 

The increased emphasis on real-time monitoring of the structures in order to detect flaws at an early stage to prevent failure led to a soar in the interest within the research community. Stochastic Bayesian methods have been established and implemented on several structural system identification problems by \cite{Beck-JL, Beck-JL2, Beck-JL3, Beck-JL4, Beck-JL5}. A Bayesian approach for quantitatively identifying damages in aluminum beam-like structures through a hybrid particle swarm optimization algorithm has been proposed in \cite{Ng-CT}. Multiple flaws in the structures have been characterized using the reversible jump Markov chain Monte-Carlo (RJMCMC) method by \cite{Yan-G}. The authors have employed extended FEM (XFEM) as the forward solver in the inverse detection framework. 
Oftentimes, the number of damages in the structure is not known prior to the characterization stage. To overcome this challenge, a Bayesian model class selection algorithm based on spectral FEM was introduced in \cite{He-S}, to determine the number of cracks in the structure without any prior knowledge. Subsequently, transitional MCMC (TMCMC) method was utilized to characterize the identified defects. In \cite{Yang-J}, the size of a crack in an aluminum plate was successfully quantified under probabilistic setting using in-situ Lamb wave testing and Bayesian updating. In the recent times, \cite{Wang-X} employed variational Bayesian inference and delayed rejection adaptive Metropolis algorithm to successfully detect damages in a two-storey model built in steel frames. The anisotropic nature of materials like composites and fiber metal laminates (FMLs) adds to the difficulty of monitoring the structural integrity. \cite{Chiachio} presented a multistage Bayesian procedure for identifying damage parameters in composite laminates using through-transmission ultrasonic measurements. \cite{Kundu-A} used a network of distributed sensors and their cross-correlation information in the Bayesian sense to localize the damage in composites. The other league of Bayesian damage identification is the time-domain approach which is appropriate with time-varying systems and where a sequential dataset is observed. Only a few investigations were performed on this category of parameter estimation routine \cite{Ching, Yuen-KV, Gao-Q, Jin-C, Mariani}. 

Although several SHM studies have been conducted on homogeneous metals and FRPs, there is, to the best of our knowledge, no work that investigated damage characterization on FMLs using GUW. In this research work, the Metropolis-Hastings algorithm centered on MCMC and Ensemble Kalman filter (EnKF) technique are exploited not only to localize and characterize the damage in the FML but also to quantify its associated uncertainties. The damage position and its size along the length of the laminate, as well as its modulus of elasticity are inferred from the noisy sensor measurements. The results corresponding to both the methods are compared and their efficiencies are critically analyzed. Both the MCMC and the EnKF method only need forward solves of the model and they provide estimates of the uncertainty of the obtained estimated parameters. In contrast, gradient-based methods for variational formulations of the inverse problem, such as the Landweber method, could be used to obtain point estimates of the parameters. While these methods would presumably need fewer forward evaluations, they would need to evaluate the gradient (which had to be derived analytically) and they would not give estimates of the uncertainty. Moreover, gradient-free solvers (i.e. metaheuristics such as genetic algorithms) could be employed. They would not need gradient information but usually need many forward solves of the model and they would not provide information about the uncertainty. Hence, we only focus on Bayesian methods in this article.

The rest of the paper is organized as follows. Section~\ref{sec:forward-problem} and Section~\ref{sec:pod-mor} describe the forward model for the system of consideration and its associated parametric model order reduction (PMOR) respectively. In Section~\ref{sec:inverse}, the inverse problem is introduced and the methodologies of MCMC and EnKF based on the Bayesian framework are presented. Following this, relevant numerical experiments are conducted and studied in Section~\ref{sec:numerical}. Finally, Section~\ref{sec:discussion} discusses the results and Section~\ref{sec:conlusion} summarizes the key findings of this research and discusses the potential scope for future work.

\section{Forward problem}
\label{sec:forward-problem}
\begin{figure}[h]
	\centering
	\includegraphics[width=15 cm]{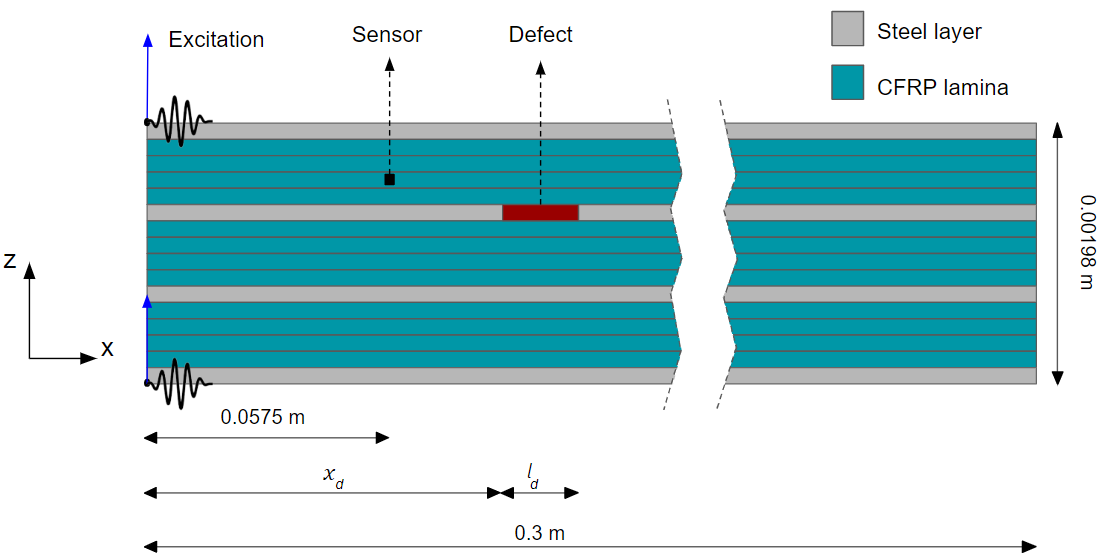}
	\caption{Two-dimensional model setup for simulation of GUW propagation in CFRP-Steel laminate.}
	\label{fig: ModelSetup}
\end{figure}

We consider the two-dimensional model of an FML as introduced in \cite{Bellam-Muralidhar-NK} summarized below, which describes the propagation of anti-symmetric Lamb wave mode ($A_0$) in an FML and its interaction with the damage. The waveguide is a sixteen-layered carbon fiber reinforced plastic (CFRP) steel laminate with a defect in one of its steel layers. The defect is modeled by locally reducing Young's modulus of the steel layer. Consequently, a reduced modulus of elasticity is applied to a certain number of elements in the defective area. The phenomena of GUW propagation and its interaction with the damage in FML is realized using COMSOL-Multiphysics software\textsuperscript{\textregistered}. The actuators are placed on the top and bottom left nodes of the model that actuates the FML with a five-cycle Hanning window sinusoidal burst with a central frequency of 120 kHz. We assume a fixed sensor location within the FML as shown in Figure \hspace{-1.5 mm}~\ref{fig: ModelSetup}. 

The time-dependent force resulted due to the excitation is represented in Figure 2(a). The simulation lasted for a total time of $2.083 \times 10^{-4}$ s with the excitation pulse longing for $4.167\times10^{-5}$ s. The thicknesses of steel and CFRP layers are 0.12 mm and 0.125 mm, respectively. 
In this model setup, we make further suppositions that only one damage exists which appears only in the steel layer and on the right-hand side of the sensor. In layered structures like composite materials and fiber metal laminates, different classes of damage can occur namely, matrix damage, fiber damage, delamination and foreign particle invasion. In this initial phase of the project, we would like to identify the damage (foreign particle invasion) causing local reduction of stiffness of a layer along the length of the laminate. In the second phase of this project, we will work with complex damages i.e. high-velocity impact damages where damage occurs across the layers. The FML boundaries are subjected to free boundary conditions. All the nodes on the left end of the FML are enforced to symmetry boundary conditions, while the center node on the right end of the laminate was fixed in the $x-$direction. We assume a fixed placement of a virtual sensor embedded at 57.5 mm along the length of the laminate that records the out-of-plane displacements during the simulation. It is supposed that the plane strain conditions be prevalent for the setting. Based on the Courant-Friedrichs-Lewy condition, the time step size is chosen to be $4.17\times 10^{-7}$ s. The discretization of the steel, CFRP layers as well as the defect of the laminate model is accomplished by two-dimensional quadrilateral plane strain elements with quadratic Lagrange shape functions resulting in a total of 79266 degrees of freedom (DOF). For further details regarding the model and material properties, we refer to Section 2 in \cite{Bellam-Muralidhar-NK}. Figure 2(b) shows the numerical simulation of the displacement field measured by the virtual sensor for the parameter $\boldsymbol{\theta} = \{4.05\textup{ GPa}, 90\textup{ mm}, 4\textup{ mm}\}$. The parameter $\boldsymbol{\theta}$, in this application is defined over a three-dimensional space by Young's modulus, position and size of the damage along the length of the laminate as $\boldsymbol{\theta}=\{E_{d}, x_{d},l_{d}\}\in\mathbb{R}^3$. We observe the excitation wave, the wave reflected from damage and its reflection from the left end of the laminate. Under the mentioned conditions and settings, the simulation time for a solve using this high-fidelity (HiFi) numerical model is 66.29 s\cite{Bellam-Muralidhar-NK}.   

\begin{figure}
	\centering
	\begin{minipage}{.5\textwidth}
		\centering
		\subfloat[]{\includegraphics[width=1\linewidth]{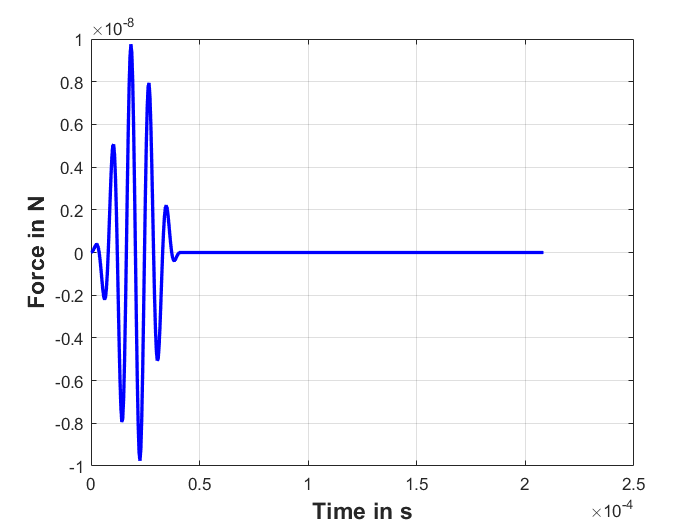}}
		\label{fig:ExcitationFunction}
	\end{minipage}%
	\begin{minipage}{.5\textwidth}
		\centering
		\subfloat[]{\includegraphics[width=1\linewidth]{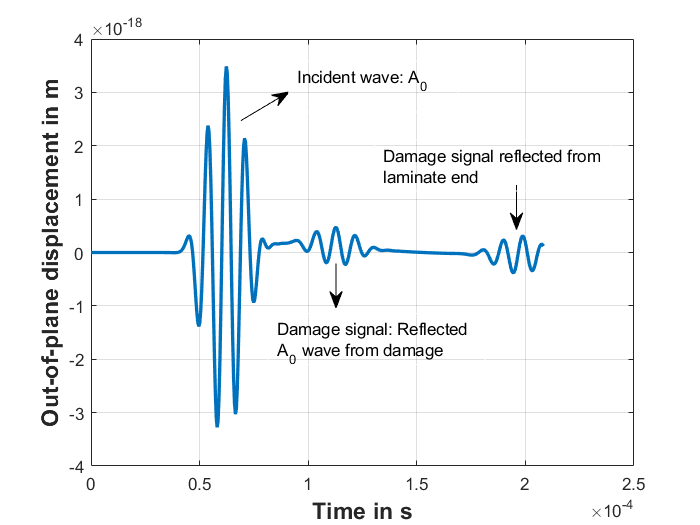}}
		\label{fig: TFol}
	\end{minipage}
	\caption{(a) Transient force applied during excitation of the FML \cite{Bellam-Muralidhar-NK} (b) Displacement measured by the sensor for parameter configuration as in Figure \hspace{-1.5 mm}~\ref{fig: ModelSetup} \cite{Bellam-Muralidhar-NK}.}
\end{figure}

\section{POD-based parametric model order reduction}
\label{sec:pod-mor}
Model order reduction (MOR) aims to reduce the huge computational cost associated with solving high-dimensional numerical models by approximating them on a lower-order space with a minimal loss in its accuracy. 
Let us first introduce the idea of MOR for an arbitrary but fixed parameter $\boldsymbol{\theta} \in \mathcal{D} \subset \mathbb{R}^{N_p}$ with $N_p$ parameters defining the system from the parametric space $\mathcal{D}$. This leads to a local ROM. Afterwards, we discuss the construction of a global ROM whose basis sufficiently well approximates the solution manifold $\mathcal{M} = \{ \mathbf{u}(\boldsymbol{\theta},t) \; | \; \mathbf{u}(\boldsymbol{\theta},t)$  solves the HiFi problem for $ \boldsymbol{\theta} \in \mathcal{D}\}$. For the introduction of the local ROM, we recall the
proper orthogonal decomposition (POD) technique, which is prominent and has been successfully applied to numerous fields viz.\ geophysical fluid dynamics, signal analysis and pattern recognition \cite{Aradag, Towne-A, Ting-Z, Prothin-S}.
Applying a finite element spatial discretization on the governing equation of the system under consideration, an undamped linear dynamic structural HiFi model results in the following system of ordinary differential equations with $N$ degrees of freedom:
\begin{eqnarray}
	\mathbf{M(\boldsymbol{\theta})\hspace{1 mm}\ddot{u}}(\boldsymbol{\theta},t) + \mathbf{K(\boldsymbol{\theta})\hspace{1 mm}u}(\boldsymbol{\theta},t) = \mathbf{f}(t) \label{eq: hifiFEM},\\
	\mathbf{u}(\boldsymbol{\theta}, t=0) = 0,\\
    \mathbf{\dot{u}}(\boldsymbol{\theta}, t=0) = 0
    \label{eq: hifiFEM2}
\end{eqnarray}
where $\mathbf{M}(\bm{\theta})\in\mathbb {R}^{N\times N}$ represents the global mass matrix, $\mathbf{K}(\bm{\theta})\in\mathbb{R}^{N\times N}$ denotes the global stiffness matrix, $\mathbf{f}(t)\in\mathbb{R}^{N}$ designates the excitation force applied and $\mathbf{u}(\boldsymbol{\theta},t)\in\mathbb{R}^{N}$ as the displacement matrix.  Let $t\in[0,t_{max}]$ represent the time variable. 
The projection of the basis functions from its high-dimensional space $\mathbb{R}^N$ to a lower-dimensional space $\mathbb{R}^n$ using a projection matrix $\mathbf{\Phi}(\boldsymbol{\theta})\in\mathbb{R}^{N\times n}$ where $n\ll N$:
\begin{equation}
	\mathbf{u}(\boldsymbol{\theta},t)\approx \mathbf{\Phi}(\boldsymbol{\theta}) 	\mathbf{u_r}(\boldsymbol{\theta},t),  \hspace{1 cm} \mathbf{\ddot{u}}(\boldsymbol{\theta},t)\approx \mathbf{\Phi}(\boldsymbol{\theta}) \mathbf{\ddot{u}_r}(\boldsymbol{\theta},t).
	\label{eq: Approximation}
\end{equation}
Substituting (\hspace{-1 mm}~\ref{eq: Approximation}) into (\hspace{-1 mm}~\ref{eq: hifiFEM}) and projecting on to the reduced space ${\mathbf{\Phi}}$ yields: 
\begin{equation}
	\underbrace{\mathbf{\Phi^T(\boldsymbol{\theta}) M(\boldsymbol{\theta}) \Phi(\boldsymbol{\theta})}}_{\mathbf{M_r}} \mathbf{\ddot{u}_r}(\boldsymbol{\theta},t) + \underbrace{\mathbf{\Phi^T}(\boldsymbol{\theta}) \mathbf{K}(\boldsymbol{\theta}) \mathbf{\Phi}(\boldsymbol{\theta})}_{\mathbf{K_r}} \mathbf{u_r}(\boldsymbol{\theta},t) = \mathbf{\Phi^T}(\boldsymbol{\theta}) \mathbf{f}(t).
\end{equation}
Introducing the reduced system matrices $\mathbf{M_r}(\boldsymbol{\theta}), \mathbf{K_r}(\boldsymbol{\theta})\in\mathbb{R}^{n\times n}$ and $\mathbf{f_r}(t)\in\mathbb{R}^{n}$ we obtain the reduced system
\begin{equation}
	\hspace{1 cm}\mathbf{M_r}(\boldsymbol{\theta}) \mathbf{\ddot{u}_r}(\boldsymbol{\theta},t) + \mathbf{K_r}(\boldsymbol{\theta}) \mathbf{u_r}(\boldsymbol{\theta},t) = \mathbf{f_r}(t).
	\label{eq: ReducedModel}
\end{equation}
The reduced-order basis (ROB) matrix $\mathbf{\Phi}(\boldsymbol{\theta})$ can be extracted using the well-known POD method. To start with, a snapshot matrix $\mathbf{S}(\boldsymbol{\theta})\in\mathbb{R}^{N\times m}$ for a particular $\boldsymbol{\theta}$ containing the high-dimensional solutions at $m$ instants of time is constructed as\begin{equation}
	\mathbf{S}(\boldsymbol{\theta}) = [\mathbf{u}(\boldsymbol{\theta},t_1), \mathbf{u}(\boldsymbol{\theta},t_2), \dots,\mathbf{u}(\boldsymbol{\theta},t_m)].
\end{equation}
In practice, we utilize the prominent Newmark's time integration method for temporal discretization and solving the system. The snapshot matrix $\mathbf{S}(\boldsymbol{\theta})$ is then decomposed using thin SVD as follows (where we omitted the dependency on the parameters $\boldsymbol{\theta}$ for readability)
\begin{equation}
	\mathbf{S} = \mathbf{W\Sigma V}^T.
	\label{eq: SVD}
\end{equation}
In ($\hspace{-1.75 mm}~\ref{eq: SVD}$), $\mathbf{W} = [\mathbf{w}_1, \mathbf{w}_2, \dots, \mathbf{w}_m]\in\mathbb{R}^{N\times m}$ is the left-singular matrix containing orthogonal basis vectors, which are also called as the proper orthogonal modes (POMs) of the system. $\mathbf{\Sigma}= \textup{diag}(\sigma_1, \sigma_2, \dots, \sigma_m)\in\mathbb{R}^{m\times m}$, with $\sigma_1 \ge \sigma_2 \ge \dots \sigma_m \geq 0$ is a diagonal matrix with singular values $\{\sigma_k\}_{k = 1}^{m}$ and $\mathbf{V}\in\mathbb{R}^{m\times m}$ represents the right-singular matrix. 

%
%

By ignoring the rest of the modes $[\mathbf{w}_{n+1}, \mathbf{w}_{n+2},\dots,\mathbf{w}_m]$, only the first $n$ number of basis vectors of the system is accumulated to form the projection matrix $\mathbf{\Phi}$ :

\begin{equation}
	\mathbf{\Phi} = [\mathbf{w}_1, \mathbf{w}_2, \dots, \mathbf{w}_n]\in\mathbb{R}^{N\times n}.
\end{equation}

As the ROB matrix $\mathbf{\Phi}$ is obtained, the reduced-order system ($\hspace{-1.5 mm}~\ref{eq: ReducedModel}$) is solved for $\mathbf{u_r}$ and $\mathbf{\ddot{u}}_r$ using the Newmark's method similar to the high-dimensional model. Eventually, the solution for the full-order system can be computed using ($\hspace{-1.5 mm}~\ref{eq: Approximation}$). The choice of the dimensions of the reduced-order model $n$ is crucial in order to accurately reproduce the dynamics of the original system. 

Often the ROMs produced by the so far discussed method lack robustness to the parameter variations. However, parameter variations have become essential in inverse problem analysis, optimization and uncertainty quantification. Therefore, it is desired to derive a global parametric ROM that is valid for all the possible values within the interested parametric domain, i.e.\ we now aim to construct a global ROM whose basis approximates the solution manifold $\mathcal{M}$ sufficiently well. This issue is precisely addressed by the parametric model order reduction approach. Most of the existing PMOR methods involve sampling the entire parametric domain and computing the HiFi solution at those sampled parameter sets. This benefits the extraction of essential global  reduced-order bases (ROBs) which could approximate the underlying high-dimensional system with great accuracy. The parameters that are sampled from the parametric domain determine the fidelity of the generated ROM. In most of the POD-based PMOR applications, the sampling of parameters is carried out in a greedy strategy - a heuristic problem-solving approach that takes the locally optimal solution at each stage that reaches the globally optimal solution over a period of time \cite{Bui-Thanh, Veroy-K, Haasdonk, Ullmann-S, Bellam-Muralidhar-NK}. At each stage, the procedure quests the parameter configuration at which the ROM yields the largest error solves the HiFi response for that parameter and thereafter updates the ROM. As the computation of exact error is impossible without evaluating the HiFi model, an error estimate is generally used. There exist several \textit{a posteriori} error estimators based on the underlying partial differential equation. Several approaches \cite{Veroy-K2, PrudHomme, Amsallem-D, Choi-Y, Khan-A} concentrate on the affine parameter dependency of the high-dimensional model resulting in an offline/online decomposition approach while some \cite{He-X-Choi, Kim-Y-Choi, Choi-Y-Carlberg, Lauzon-J, Negri-F} do not rely on it.


An  approach based on the concept of optimization techniques that sample parameters in an adaptive manner and efficiently draws out the global ROBs is employed in this research work. The readers are directed to \cite{Bellam-Muralidhar-NK} for a detailed description of the development and analysis of the PMOR for the system. The proposed procedure also employs a surrogate model which is constructed upon the evaluated \textit{a posteriori} error indicators, the norm of the residual corresponding to the system, that enables a swift localization of the parametric domain corresponding to the highest probability of error. The proposed method was successfully applied to the system considered, GUW propagation in an FML structure containing a homogeneous defect, and an efficient global ROM was developed. The training parametric space defined as $\mathcal{D} = \{(E_d, x_d, l_d)\ |\ E_d\in[E_{\min},E_{\max}]\textup{ Pa},\hspace{2 mm} x_d\in[x_{\min},x_{\max}]\textup{ mm}, \hspace{2 mm} l_d\in[l_{\min},l_{\max}]\textup{ mm}\}$ is bounded by $E_{d}\in [5\times10^6, 5\times10^9]$ Pa, $x_{d}\in [20, 150]$ mm and $l_{d}\in [2, 15]$ mm. The subscript $d$ refers to damage. 

\begin{figure}[H]
	\centering
	\includegraphics[width=13 cm]{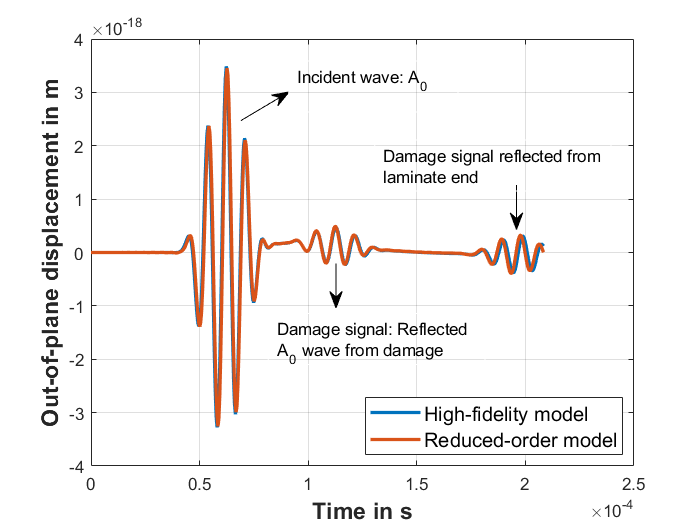}
	\caption{Comparison of reduced-order solution with the HiFi solution for the parameter configuration as in Figure~\ref{fig: ModelSetup} \cite{Bellam-Muralidhar-NK}.}
	\label{fig: HDMvsROS}
\end{figure}

\begin{figure}[H]
	\centering
	\begin{minipage}{.5\textwidth}
		\centering
		\subfloat[]{\includegraphics[width=1\linewidth]{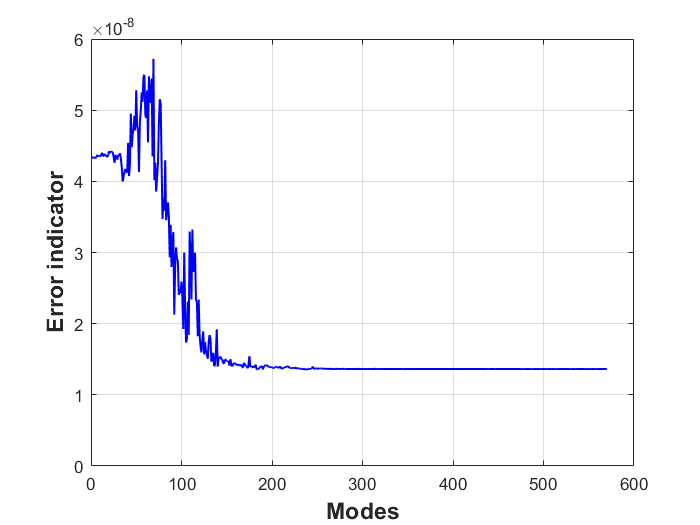}}
		\label{fig: EI}
	\end{minipage}%
	\begin{minipage}{.5\textwidth}
		\centering
		\subfloat[]{\includegraphics[width=1\linewidth]{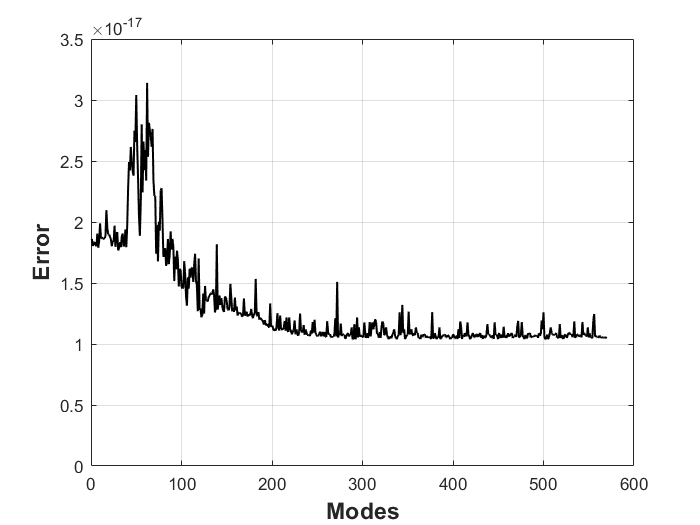}}
		\label{fig: RE}
	\end{minipage}
	\caption{ Evolution of (a) error indicator and (b) error in the measurement of out-of-plane displacement between the HiFi solution and the reduced-order solution \cite{Bellam-Muralidhar-NK}.}
	\label{fig: Evolutionplots}
\end{figure}

Using the developed PMOR, the evaluation of the displacement field for any given parameter configuration is accelerated by a factor of 33.82. Figure \hspace{-1.5 mm}~\ref{fig: HDMvsROS} represents the comparison plot of displacement computed by HiFi model and global ROM obtained by PMOR. It can be observed that the reduced-order solution accurately approximates its HiFi counterpart with a minor loss of information. This level of accuracy was achieved with 340 global basis functions that were accumulated in a greedy fashion. The evolution plots of error indicator as well as the error between the HiFi model and the developed ROM during the adaptive greedy procedure presented in Figure \hspace{-1.5 mm}~\ref{fig: Evolutionplots}, justifies the chosen number of modes. The error here refers to the Euclidean norm of the difference between the displacement corresponding to HiFi model and ROM, measured at the sensor position. 

\section{Inverse problem}
\label{sec:inverse}
\subsection{Overview}
Inverse problems probe into unknown causes or parameters of the observed data, based on predictive models, usually referred to as forward models, that relate the former quantities to the latter in the causal order. They are evident in several applications like system identification, optics, acoustics, medical imaging, computer vision and structural health monitoring. Inverse problems are generally ill-posed, problems that satisfy at least one of the following conditions: a) no existence of solution, b) insufficient data to estimate the parameters unambiguously and c) a small perturbation due to measurement noise and modeling errors in the data would lead to large variations in the parameter estimation, i.e., Hadamard instability. The characterization of ill-posedness can be associated with the demonstration of the Second Law of Thermodynamics: the entropy of an irreversible system increases as it loses information inevitably \cite{Calvetti}. Therefore, estimating the parameters of the system by employing noisy measurement data and predictive models poses several non-trivial challenges. 


Inverse problems are conventionally approached from the viewpoint of regularization, the process of trading an unstable inverse problem with a nearly stable one. This approach results in a specific value of parameters of the system. However, the measurement data from the experiments in the laboratory or field are usually prone to systematic errors due to the inaccuracy of measurement devices or because of human factors. As a consequence, the point estimates of the parameters are often uncertain and could possibly be away from the true value. In contrast to the regularization techniques, the Bayesian framework provides new methods to handle the inaccuracies and inadequacies of the forward problem in the stochastic sense. In a Bayesian inference method, the unknown parameter is modeled as a random variable and is described by means of its distribution rather than a specific value. This method also has a provision for incorporating prior knowledge about the possible values for the uncertain parameters of the system with the information obtained from the field or experiments. This could, in general, reduce the uncertainty concerning the parameter of interest and result in improved models. In a stochastic setting, the assimilation of new information obtained through measurement into the prior model is accomplished by conditioning the model on available information. This results in a conditional measure, based on Kolmogorov's concept of conditional expectation. In particular, Bayes' formula provides an analytical expression for the conditional measure and serves as the main tool in Bayesian inference. 


%
As SHM, in general, seeks to identify the damage with a minimal number of sensors $s$, the field variable is not monitored at all the available spatial grid points. Therefore, an input-output map with an observation operator is considered:
\begin{equation}
	\boldsymbol{\theta} \mapsto \mathbf{s}_h(\boldsymbol{\theta}) = \mathbf{B u}(\boldsymbol{\theta}) \in \mathbb{R}^{s\times m}
	\label{eq: ipopMap}
\end{equation}
where $\mathbf{B}\in\mathbb{R}^{s\times N}$ is the observation matrix, $\mathbf{u}(\boldsymbol{\theta})\in\mathbb{R}^{N\times m}$ represents the solution of the system containing the state of the system given by $N$ spatial degrees of freedom and at $m$ time points, and $\mathbf{s}_h(\boldsymbol{\theta})$ is the observation of the system. In this research work, only one sensor is used to monitor the out-of-plane displacement of the laminate and hence, $s=1$. An additive noise for the observed quantity is assumed: $\mathbf{s}(\boldsymbol{\theta}) = \mathbf{s}_h(\boldsymbol{\theta}) + \varepsilon$, with the noise being independent between the time steps and having a Gaussian distribution, $\varepsilon\sim\mathcal{N}(0,\sigma^2)$. 

Assuming $\pi_{prior}(\boldsymbol{\theta})$ denotes the prior probability density of the unknown parameter and $\pi(\mathbf{s}|\boldsymbol{\theta})$ representing the likelihood density, the belief is updated using the Bayes' rule:

\begin{equation}
	\pi_{post}(\boldsymbol{\theta}|\mathbf{s}) = \frac{\pi(\mathbf{s}|\boldsymbol{\theta}) \hspace{1.5 mm} \pi_{prior}(\boldsymbol{\theta})}{\pi(\mathbf{s})}.
	\label{eq: BayesRule}
\end{equation}

The posterior density $\pi_{post}(\boldsymbol{\theta}|\mathbf{s})$ in (\hspace{-1.5 mm}~\ref{eq: BayesRule}) constitutes the complete solution of the inverse solution in the Bayesian framework. The denominator term $\pi(\mathbf{s})$ is called as the density of marginal likelihood or evidence of the data and is defined as,
\begin{equation}
	\pi(\mathbf{s}) = \int_{\mathbb{R}^n} \pi(\mathbf{s}|\boldsymbol{\theta}) \pi_{prior}(\boldsymbol{\theta}) d\boldsymbol{\theta}
	\label{eq: MarginalDensity}
\end{equation}
and is usually ignored in most of the Bayesian inference methods owing to two reasons: 1) as it just serves the purpose of normalization and 2) the computational effort is exorbitantly high when the number of parameters is large. Often the targeted posterior distribution is only known up to its proportionality:
\begin{equation}
	\pi_{post}(\boldsymbol{\theta}|\mathbf{s}) \propto P_{post}(\boldsymbol{\theta}|\mathbf{s}) = \pi(\mathbf{s}|\boldsymbol{\theta}) \pi_{prior}(\boldsymbol{\theta}).
	\label{eq: PropBayes}
\end{equation}

Due to the growing available computing resources as well as efficient numerical methods for high-dimensional problems, the interest and research activity in uncertainty quantification for several sophisticated systems has surged significantly. Consequently, the Bayesian inverse problems have gathered humongous attention within the community of scientific computing which lead to the development of several algorithms for their solution. Of the existing methods, in this research paper, the central attention is given to the application of Markov Chain Monte-Carlo (MCMC) based Metropolis-Hastings algorithm \cite{Metropolis, Hastings} and Ensemble Kalman Filter technique (EnKF) \cite{Evensen} to identify the damage present in the FML.
The Metropolis-Hasting method uses the full parameter to state map to update the posterior of the parameters after every full forward solve of the model, while the EnKF method propagates a so-called ensemble consisting of several parameters sets forward in time simultaneously and updates the values of all parameter sets in the ensemble every few time steps.

\subsection{Markov chain Monte-Carlo}

Markov chain Monte-Carlo is a technique employed for estimating the unknown quantity of interest using the random walk approach. It consists of a class of algorithms for successive sampling from a probability distribution based on constructing a Markov chain whose equilibrium distribution is its target distribution. The general notion of MCMC is to compute the probabilities of the samples drawn from the Markov chain and average them to estimate the solution. As the name indicates, MCMC is a unification of two approaches: Markov chain and Monte-Carlo. 

A discrete-time stochastic process $X = \{X_\zeta : \zeta \ge 0\}$ on a finite set $S$ is said to be a Markov chain if it satisfies the Markov property. The set $S$ is the state space of the process and $X_\zeta\in S$ is the state of the process at time $\zeta$. For any $i, j \in S$ and $\zeta \ge 0$,
\begin{equation}
	P\{X_{\zeta+1} = j | X_0, X_1, \dots, X_\zeta\} = P\{X_{\zeta+1} = j | X_\zeta\}, \label{eq: Mchain_a}
\end{equation}
\begin{equation}
	\hspace{2.3 cm}P\{X_{\zeta+1} = j | X_\zeta = i\} = p_{ij}. 
	\label{eq: Mchain_b}
\end{equation}
The Markov property (\ref{eq: Mchain_a}), illustrates that the next state of the process $X_{\zeta+1}$ conditionally depends only on the current state $X_\zeta$ and is independent on the previous states $X_0, X_1, \dots, X_{\zeta-1}$. The probability that the Markov chain moves from state $i$ to state $j$, represented by $p_{ij}$, is referred to as transition probability. The sum of these transition probabilities in the state space of the process is equal to one, i.e. $\sum_{j\in S} p_{ij} = 1$, $i\in S$. The matrix containing these transition probabilities is called the transition matrix of the chain. The Markov property is an elementary condition that is satisfied by the state of many stochastic phenomena. The Markov chain is time-homogeneous, which is evident from (\hspace{-2 mm}~\ref{eq: Mchain_b}), as the transition probabilities are independent of time parameter $\zeta$. 

Monte Carlo is the method of computing the moments of distribution by extracting random samples from the distribution. For instance, the computation of the expectation of a real-valued function $g(X)$ on the state space:
\begin{equation}
	\mu = \mathbb{E}\{g(X)\}
	\label{eq: Mean}
\end{equation}
using exact analytical methods can be replaced by simulating for $X_0, X_1, \dots, X_\zeta$ and computing their mean as follows:
\begin{equation}
	\tilde{\mu} = \frac{1}{\zeta} \sum_{i=1}^{\zeta} g(X_i).
	\label{eq: A_Mean}
\end{equation}
The expectation $\mu$ is approximated by $\tilde{\mu}$ with $X_1, X_2, \dots, X_\zeta$ which are independent and identically distributed (IID) as $X$. In a similar fashion, the Monte-Carlo approximated variance $\tilde{\sigma}^2$ can also be estimated as:

\begin{equation}
	\tilde{\sigma}^2 = \frac{1}{\zeta} \sum_{i=1}^{\zeta} (g(X_i) - \tilde{\mu})^2.
	\label{eq: A_variance}
\end{equation}

MCMC methods are essentially used for deriving numerical approximations of high-dimensional integrals, especially in Bayesian inference, where the analytical solution of the posterior distribution is often practically intractable. There are several approaches to implement MCMC in order to successfully approximate the targeted posterior distributions of parameters in a Bayesian MCMC model. Some of them are Metropolis-Hastings algorithm, Gibbs sampler \cite{Geman}, Slice sampler \cite{Slice}, Reversible-Jump MCMC \cite{Green} and Hamiltonian Monte-Carlo \cite{Duane-S}. In this research work, the most popular MCMC method, Metropolis-Hasting (MH) algorithm, was employed to identify the parameters of the damage in the laminate. It is a random walk procedure that utilizes an acceptance/rejection rule to converge to the target distribution $\pi_{post}(\boldsymbol{\theta}|\mathbf{s})$. The algorithm requires the choice of the proposal distribution $Q(\boldsymbol{\theta}^*|\boldsymbol{\theta})$ from which the candidate sample, $\boldsymbol{\theta}^*$ for the new Markov chain state is drawn. The MH algorithm is summarized in Algorithm \hspace{-1 mm}~\ref{algo1}:

\begin{algorithm}
	\textbf{Input:} Proposal distribution $Q(\boldsymbol{\theta})$, maximum number of samples $\Lambda$ to be drawn, target distribution $P(\boldsymbol{\theta})$
	
	\textbf{Output:} Samples from the targeted posterior distribution
	\caption{Metropolis-Hastings algorithm}
	\begin{algorithmic}[1]
		
		\STATE Select an initial value for the parameter, $\boldsymbol{\theta}_0$ based on the prior knowledge on the parameter
		
		\STATE \textbf{for} $i = 1, \dots, \Lambda$ \textbf{do}
		
		\STATE \hspace{0.5 cm} Draw a candidate sample for the parameter, $\boldsymbol{\theta}^*$ from the proposal distribution\\ \indent \hspace{5 mm} given the value of the parameter in previous iteration: $\boldsymbol{\theta}^* \sim Q(\boldsymbol{\theta}^*|\boldsymbol{\theta}_{i-1})$  
		
		\STATE \hspace{0.5 cm} Compute the probability of accepting the candidate drawn from the proposal\\ \indent \hspace{5 mm}  distribution by using Metropolis-Hastings criterion:
		\vspace{3 mm}
		
		\begin{equation}
			\alpha =  \frac{P(\boldsymbol{\theta}^*)Q(\boldsymbol{\theta}_{i-1}|\boldsymbol{\theta}^*)}{P(\boldsymbol{\theta}_{i-1}) Q(\boldsymbol{\theta}^*|\boldsymbol{\theta}_{i-1})} 
		\end{equation}
		\vspace{3 mm}
		
		\STATE \hspace{0.5 cm} Draw a sample $\beta$ from a uniform distribution on $[0,1]$: $\beta \sim U(0,1)$
		
		\STATE \hspace{0.5 cm} Accept or reject the candidate $\boldsymbol{\theta}^*$ based on the following condition: 
		
		\hspace{0.5 cm} \textbf{if} $\alpha > \beta$ \textbf{then}
		
		\hspace{1 cm} Accept $\boldsymbol{\theta}^*$ and move to the new state by setting $\boldsymbol{\theta}_i = \boldsymbol{\theta}^*$
		
		\hspace{0.5 cm} \textbf{else} 
		
		\hspace{1 cm} Reject $\boldsymbol{\theta}^*$ and remain in the same state by setting $\boldsymbol{\theta}_i = \boldsymbol{\theta}_{i-1}$
		
		\STATE \textbf{end for}
	\end{algorithmic}
	\label{algo1}
\end{algorithm}

The selected proposal distribution must be able to generate samples belonging to the support of the target distribution. If the proposal distribution is symmetric, $Q(\boldsymbol{\theta}^*|\boldsymbol{\theta}_{i-1}) = Q(\boldsymbol{\theta}_{i-1}|\boldsymbol{\theta}^*)$, then the probability of moving to the new state reduces to $P(\boldsymbol{\theta}^*)/P(\boldsymbol{\theta}_{i-1})$; therefore, if $P(\boldsymbol{\theta}^*) \ge P(\boldsymbol{\theta}_{i-1})$, the chain definitely moves to $\boldsymbol{\theta}^*$, else it moves with the probability $\alpha$.

\subsection{Ensemble Kalman Filter}

Ensemble Kalman filter (EnKF) is a data assimilation technique that combines the Kalman filter (KF) with a suitable sampling strategy. In contrast to the KF, the EnKF stores, propagates and sequentially updates through time a finite ensemble of vectors that approximate the distribution of parameters rather than updating a single parameter. This is also a sort of dimension reduction, as only a small ensemble is propagated in lieu of the joint distribution which includes the full covariance matrix. As and when the new observations are available, the parameter ensemble is updated based on the assumption of linearity and Gaussianity of the model by comparing the state produced by the system with the measurements at the current time point. Therefore, additional modifications of smoothers are required when non-linearity and non-Gaussianity are involved in the model. The use of a linear updating rule that converts the prior ensemble to a posterior ensemble after each observation inflow in EnKF makes it different from the other sequential Monte-Carlo algorithms (e.g., Metropolis-Hastings). 

%
%
As the parameter estimation procedure in EnKF is carried out successively over the interval $t\in(0,T)$, the input-output map ($\hspace{-1.5 mm}~\ref{eq: ipopMap}$) should be evaluated at several time instants $k$, which can be different to $j$, at which the underlying dynamical system is solved \cite{Pagani}. The time interval $(0,T)$ is discretized into $N_\tau$ number of time windows $(\tau^{k}, \tau^{k+1})$ of length $\Delta \tau = K\Delta t$, with $k = 0, 1, 2, \dots, N_{\tau}-1$ with $K>1$ (see Figure 5). The frequency of the EnKF update $K$, can be crucial in order to obtain an accurate estimation of the model parameters. 

\begin{figure}[h]
	\centering
	\includegraphics[width=13 cm]{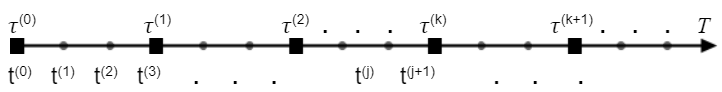}
	\caption{An example of a time grid for the EnKF update with windows of length $\Delta \tau= K\Delta t$ with $K = 3$.}
	\label{fig:TimeInterval}
\end{figure}

The update of a particle ensemble $\mathcal{P}^{(k)}$, a sample of $N_e$ parameter vectors:

\begin{equation}
	\mathcal{P}^{(k)} = \{\boldsymbol{\theta}^{(k)}_{\rho}\}_{\rho = 1}^{N_e}, \hspace{5 mm} k = 1, 2, \dots, N_\tau
	\label{eq: particleEnsemble}
\end{equation}

is accomplished by the prediction-analysis procedure. In (\hspace{-1.5 mm}~\ref{eq: particleEnsemble}), $\boldsymbol{\theta}^{(k)}_{\rho}$ denotes value of a parameter vector $\boldsymbol{\theta}_{\rho}$ at $k^{th}$ iteration of the EnKF procedure. Let
\begin{equation}
	\hspace{-10 mm} \bar{\mathbf{u}}^{(k)} = \frac{1}{N_e} \sum_{\boldsymbol{\theta}\in\mathcal{P}^{(k-1)}} \mathbf{u}^{(k)}(\boldsymbol{\theta}), \hspace{0.2 cm} 
	\bar{\mathbf{s}}^{(k)} = \frac{1}{N_e} \sum_{\boldsymbol{\theta}\in\mathcal{P}^{(k-1)}} \mathbf{s}^{(k)}(\boldsymbol{\theta}), \hspace{0.2 cm} 
	\bar{\boldsymbol{\theta}}^{(k)} = \frac{1}{N_e} \sum_{\boldsymbol{\theta}\in\mathcal{P}^{(k)}}\boldsymbol{\theta}
	\label{eq: Means}
\end{equation}
be the sample mean vectors of the ensemble at any given $k = 1, 2, \dots, N_\tau$. Similarly, the sample covariance of the outputs and their cross-covariances are obtained in the following manner:
\begin{equation}
	C_{\mathbf{s s}}^{(k)} = \frac{1}{(N_e-1)} \sum_{\boldsymbol{\theta}\in\mathcal{P}^{(k-1)}}(\mathbf{s}^{(k)}(\boldsymbol{\theta}) - \bar{\mathbf{s}}^{(k)}) (\mathbf{s}^{(k)}(\boldsymbol{\theta}) - \bar{\mathbf{s}}^{(k)})^T 
	\hspace{2 mm} \in \mathbb{R}^{s\times s},
	\label{eq: Covariance1}
\end{equation}
\begin{equation}
	C_{\boldsymbol{\theta} \mathbf{s}}^{(k)} = \frac{1}{(N_e-1)}\sum_{\boldsymbol{\theta}\in\mathcal{P}^{(k-1)}}(\boldsymbol{\theta}^{(k)} - \bar{\boldsymbol{\theta}}^{(k)})  (\mathbf{s}^{(k)}(\boldsymbol{\theta}) - \bar{\mathbf{s}}^{(k)})^T
	\hspace{2 mm} \in \mathbb{R}^{N_p\times s},
	\label{eq: Covariance2}
\end{equation}
\begin{equation}
	C_{\mathbf{u s}}^{(k)} = \frac{1}{(N_e-1)}\sum_{\boldsymbol{\theta}\in\mathcal{P}^{(k-1)}}(\mathbf{u}^{(k)}(\boldsymbol{\theta}) - \bar{\mathbf{u}}^{(k)}) (\mathbf{s}^{(k)}(\boldsymbol{\theta}) - \bar{\mathbf{s}}^{(k)})^T
	\hspace{2 mm} \in \mathbb{R}^{N\times s}.
	\label{eq: Covariance3}
      \end{equation}
      
After initializing the particle ensemble $\mathcal{P}^{(0)}$ for $k = 0$ based on the prior knowledge regarding the unknown quantity of interest, the prediction-analysis procedure of the EnKF is executed by a two-phase process:
\begin{enumerate}
	\item \textbf{Prediction phase}: 
	
	\begin{enumerate}
		\item Evaluate the solution $\mathbf{u}^{(k+1)}(\boldsymbol{\theta})$ of the system using the ROM ($\hspace{-1.5 mm}~\ref{eq: ReducedModel}$) and ($\hspace{-1.5 mm}~\ref{eq: Approximation}$) for every particle $\boldsymbol{\theta}\in\mathcal{P}^{(k)}$ and derive their corresponding observation $\mathbf{s}^{(k+1)}(\boldsymbol{\theta})$.
		
		\item Subsequently, compute means of the sample: $\bar{\mathbf{u}}^{(k+1)}$, $\bar{\mathbf{s}}^{(k+1)}$ and $\bar{\boldsymbol{\theta}}^{(k+1)}$.
		
		\item Then compute the sample covariances: $C_{\mathbf{ss}}^{(k+1)}$, $C_{\boldsymbol{\theta}\mathbf{s}}^{(k+1)}$ and $C_{\mathbf{us}}^{(k+1)}$.
	\end{enumerate} 

	\item \textbf{Analysis phase}:
	
	Exploiting the new information from the prediction phase, update the parameter ensemble using KF updating formula:
	
	\begin{equation}
		\boldsymbol{\theta}_{\rho}^{(k+1)} = \boldsymbol{\theta}_{\rho}^{(k)} + C_{\boldsymbol{\theta}\mathbf{s}}^{(k+1)} \{(\sigma^2 + C_{\mathbf{ss}}^{(k+1)})^{-1} (\mathbf{s}_q^{(k+1)}(\boldsymbol{\theta}) - \bar{\mathbf{s}}^{(k+1)}(\boldsymbol{\theta}_{\rho}^{(k)}))\}
		\label{eq: analysisStage}
	\end{equation} 

	for all ensemble members $\rho = 1,\dots,N_e$, where $\mathbf{s}_q$ is sensor observation concerning the parameter to be identified $\boldsymbol{\theta}_{true}$. In this procedure, we do not update the state of the system but only the parameters in the ensemble.  
\end{enumerate} 

The procedure of EnKF update is as outlined in algorithm 2. 

\begin{algorithm}
	\textbf{Input:} Number of parameters in an ensemble $N_e$, Frequency of update $K$, Parametric space $\mathcal{D}$, Predictive forward ROM, Noisy observation $\mathbf{s}_q$ corresponding to the parameter $\boldsymbol{\theta}_{true}$ to be identified
	
	\textbf{Output:} Updated particle ensemble vector
	\caption{Ensemble Kalman Filter procedure}
	\begin{algorithmic}[1]
		
		\STATE Initialize the particle ensemble $\mathcal{P}^{(0)}$ based on the prior knowledge of the unknown parameters
		
		\STATE \textit{Prediction-Analysis stage}:
		
		\STATE \textbf{for} $ k = 0, \dots, N_\tau-1$ \textbf{do}
		
		\STATE \hspace{0.5 cm} \textit{Prediction stage}:
		
		\STATE \hspace{0.5 cm} \textbf{for} $\boldsymbol{\theta}^{(k)}\in\mathcal{P}^{(k)}$ \textbf{do}

		\STATE \hspace{1 cm} Compute the solution $\mathbf{u}(\boldsymbol{\theta})$ and observation $\mathbf{s}(\boldsymbol{\theta})$ for every parameter in the\\ \indent \hspace{1 cm}  ensemble using the ROM (\hspace{-1.5 mm}~\ref{eq: ReducedModel}) and (\hspace{-1.5 mm}~\ref{eq: Approximation}) 
		
		\STATE \hspace{0.5 cm} \textbf{end for}
		
		\STATE \hspace{0.5 cm} Compute the sample means: $\bar{\mathbf{u}}^{(k+1)}$, $\bar{\mathbf{s}}^{(k+1)}$ and $\bar{\boldsymbol{\theta}}^{(k+1)}$ by (\hspace{-1.5 mm}~\ref{eq: Means})
		
		\hspace{0.5 cm} Compute the sample output covariances and cross-covariances: $C_{\mathbf{ss}}^{(k+1)}$, $C_{\boldsymbol{\theta}\mathbf{s}}^{(k+1)}$ \\ \indent \hspace{0.5 cm} and $C_{\mathbf{us}}^{(k+1)}$ by (\hspace{-1.5 mm}~\ref{eq: Covariance1}), (\hspace{-1.5 mm}~\ref{eq: Covariance2}), and (\hspace{-1.5 mm}~\ref{eq: Covariance3}), respectively
		
		\STATE \hspace{0.5 cm} \textit{Analysis stage}:
		
		\STATE \hspace{0.5 cm} \textbf{for} $\boldsymbol{\theta}^{(k)}\in\mathcal{P}^{(k)}$ \textbf{do}
		
		\STATE \hspace{1 cm} Update the parameter ensemble using the KF updating formula as in (\hspace{-1.5 mm}~\ref{eq: analysisStage})
		
		\STATE \hspace{0.5 cm} \textbf{end for}
		
		\STATE \textbf{end for}
	\end{algorithmic}
	\label{algo2}
\end{algorithm}

\section{Numerical experiments}
\label{sec:numerical}
\begin{figure}[H]
	\centering
	\begin{minipage}{.5\textwidth}
		\centering
		\subfloat[]{\includegraphics[width=0.8\linewidth]{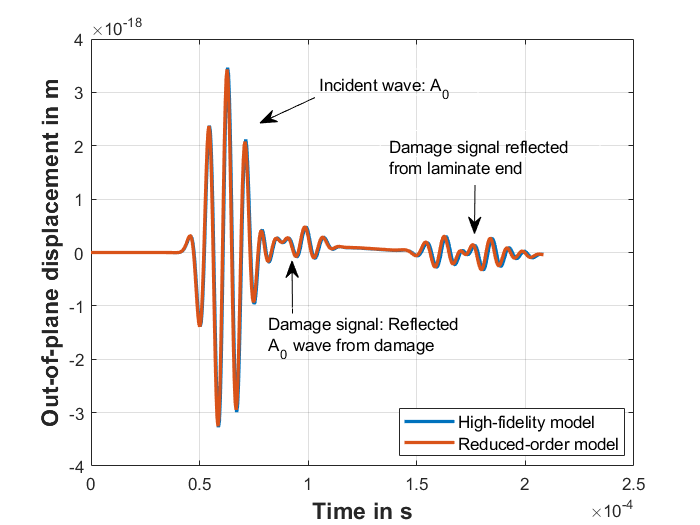}}
		\label{fig:HDM_ROM_2}
	\end{minipage}%
	\begin{minipage}{.5\textwidth}
		\centering
		\subfloat[]{\includegraphics[width=0.8\linewidth]{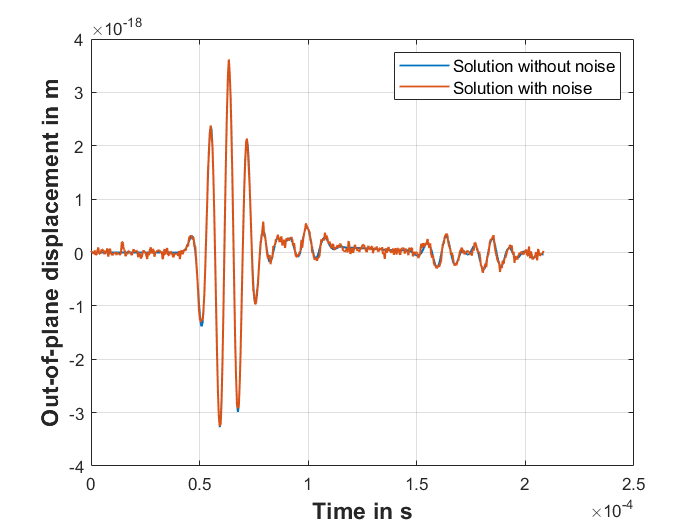}}
		\label{fig:WithandWithoutnoise}
	\end{minipage}
	\caption{(a) Accuracy of the reduced-order solution for $\boldsymbol{\theta}_{true}$  (b) Noisy signal observed by the structure embedded sensor.}
	\label{fig: NoisySignals}
\end{figure}

In this section, the performance of the proposed Bayesian methodologies: MCMC-MH algorithm and EnKF technique to estimate the damage parameters in FML structure are analyzed through numerical experiments using COMSOL-Multiphysics$\textsuperscript{\textregistered}$ and \MATLAB (the \MATLAB code is available at \url{https://github.com/nandakishorebm1995/FOR3022_Code}). We recall that the HiFi model of the underlying system consisting of 79266 DOFs is approximated onto the reduced space formed by 340 global ROB functions over the trained parametric domain $\mathcal{D}$. As the ROM is used for the forward solve, it is assumed that the damage characteristics lie within $\mathcal{D}$, i.e. $\boldsymbol{\theta}\in\mathcal{D}$, in regard to the ROM (see Section 5 in \cite{Bellam-Muralidhar-NK}). In a real deployable structural health monitoring system, the sensing elements will be affected by noise that is dependent on several factors like built-up electronic units and environmental conditions. This could crank up the difficulty to parameterize the prior in the Bayesian approach. However, at this initial phase of the research, it is assumed that the sensor is affected only by Gaussian noise, an assumption that has been confirmed by the project partners who manufacture the sensors. We further suppose a deterministic model and postpone our investigation on accounting for the model uncertainty in our future studies. The measurement data required for estimating the damage parameters is synthetically produced by adding Gaussian noise to the model observation of a certain parameter configuration that we seek to identify. The additive noise $\varepsilon$ is sampled from zero-mean Gaussian distribution: $\varepsilon\sim \mathcal{N}(0,\sigma^2)$. In this study, a $10\hspace{1 mm} \%$ of the maximum out-of-plane displacement of the signal reflected from the damage $u_{d,\max}$, is considered as the standard deviation $\sigma$. Both MCMC-MH as well as EnKF aims to identify the damage with its parameter configuration $\boldsymbol{\theta}_{true} = \{0.8 \textup{ GPa}, 70 \textup{ mm}, 10 \textup{ mm}\}$, using the noisy measurement data and the predictive forward ROM. Figure \hspace{-1.5 mm}~\ref{fig: NoisySignals}(a) depicts the accuracy of the adapted PMOR technique in approximating the underlying HiFi model for the considered parameter $\boldsymbol{\theta}_{true}$. The noisy observation of the system by the integrated sensor is shown in Figure \hspace{-1.5 mm}~\ref{fig: NoisySignals}(b). 

\vspace{-0.3 cm}
\subsection{MCMC-MH algorithm} 

\begin{figure}[H]
	\centering
	\begin{minipage}{.33\textwidth}
		\centering
		\subfloat[]{\includegraphics[width=5.5 cm,height=4.8 cm]{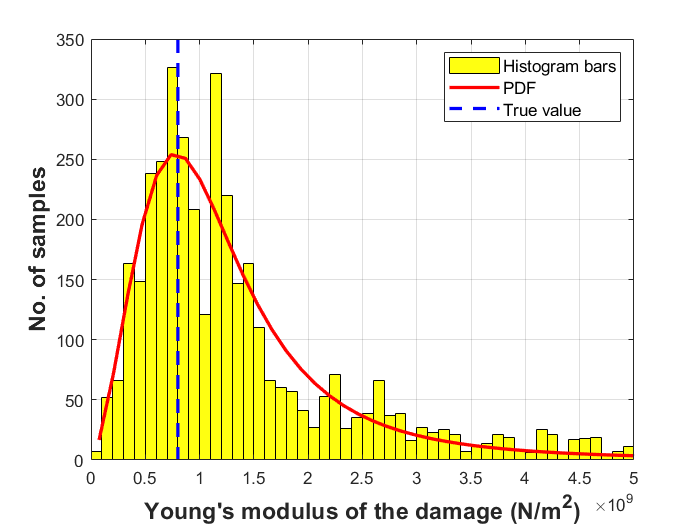}}
		\label{fig: StiffnessHist}
	\end{minipage}%
	\begin{minipage}{.33\textwidth}
		\centering
		\subfloat[]{\includegraphics[width=5.5 cm,height=4.8 cm]{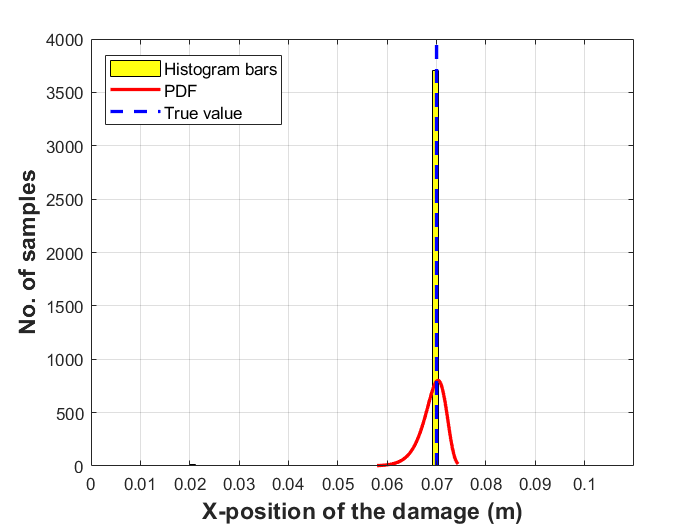}}
		\label{fig: XposHist}
	\end{minipage}
	\begin{minipage}{.33\textwidth}
		\centering
		\subfloat[]{\includegraphics[width=5.5 cm,height=4.8 cm]{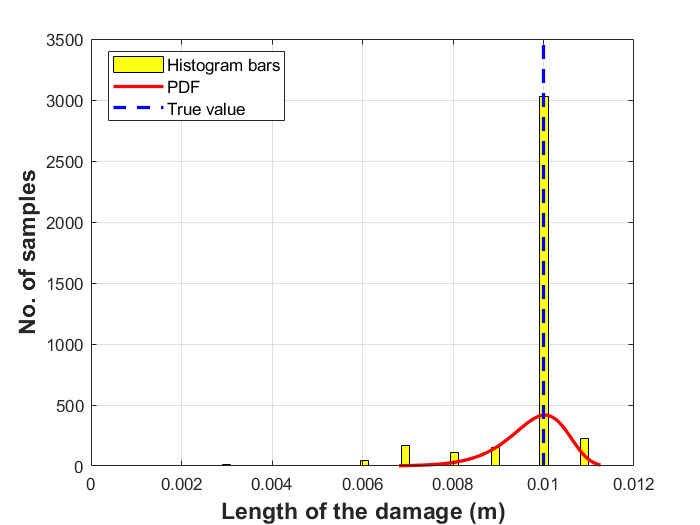}}
		\label{fig: LengthHist}
	\end{minipage}
	\caption{Histogram of the samples drawn from the posterior distribution of (a) elasticity modulus of the damage (b) position of the damage along the length of the FML structure (c) length of the damage in the structure, for $10\%$ noise level.}
	\label{fig: Histograms1}
\end{figure}

Initially, the Bayesian inference using MCMC-MH sampling is studied to identify the damage. The initial values of the prior distribution concerning the parameters are chosen in a completely random fashion from the trained parametric space $\mathcal{D}$. Following the procedure described in Section 4.2, the histogram of samples drawn from the posterior distribution of damage parameters is shown in Figure \hspace{-1.5 mm}~\ref{fig: Histograms1}. It can be observed that the parameters, position and length of the damage, are quite certainly inferred while the elasticity modulus of the damage has some uncertainty associated. However, the peak of its probability density function (PDF) is in good agreement with the true value of Young's modulus of the damage.  

\begin{figure}[h]
	\centering
	\begin{minipage}{.33\textwidth}
		\centering
		\subfloat[]{\includegraphics[width=5.5 cm,height=4.8 cm]{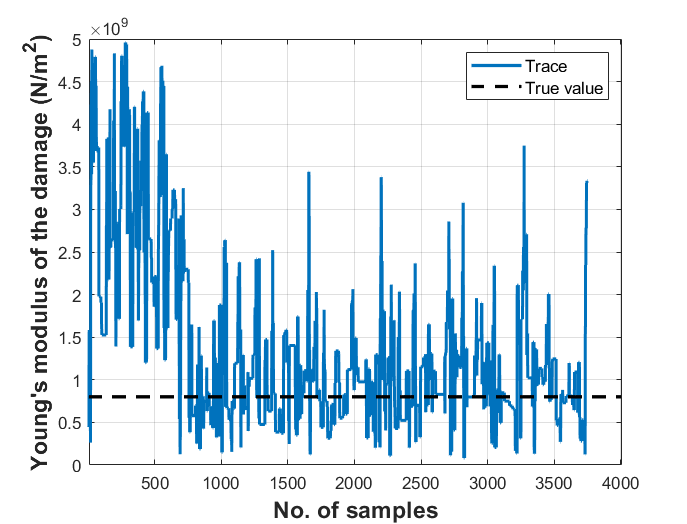}}
		\label{fig: StiffnessTace1}
	\end{minipage}%
	\begin{minipage}{.33\textwidth}
		\centering
		\subfloat[]{\includegraphics[width=5.5 cm,height=4.8 cm]{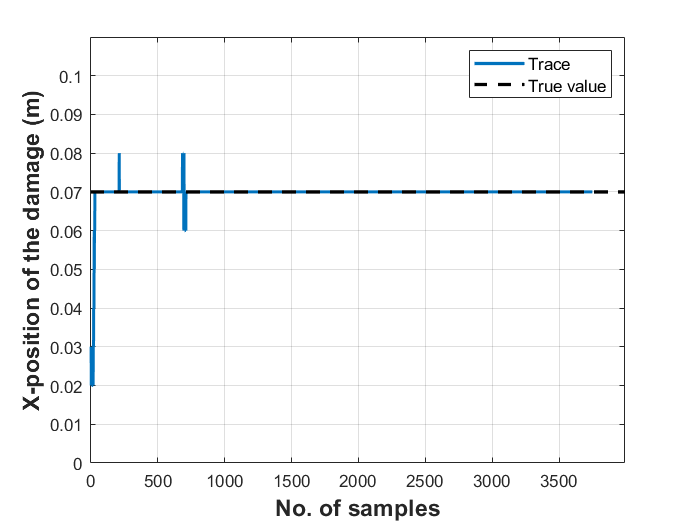}}
		\label{fig: XposTace1}
	\end{minipage}
	\begin{minipage}{.33\textwidth}
		\centering
		\subfloat[]{\includegraphics[width=5.5 cm,height=4.8 cm]{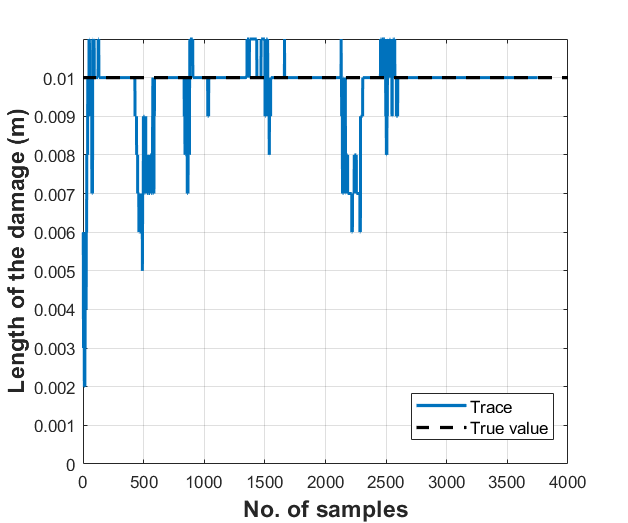}}
		\label{fig: LengthTace1}
	\end{minipage}
	\caption{Trace of MCMC-MH sampling procedure: (a) elasticity modulus of the damage (b) position of the damage along the length of the FML structure (c) length of the damage in the structure.}
	\label{fig: Trace1}
\end{figure}

The trace of the parameter sampling obtained during the MCMC-MH procedure can be seen in Figure \hspace{-1.5 mm}~\ref{fig: Trace1}. After the "burn-in" phase, i.e. the samples required for the Markov chain to reach its stationarity, of approximately 600 samples, the sampling trace of Young's modulus of the damage steers towards its true value. Totally, 3750 samples are drawn from the posterior distribution to reliably infer all the damage parameters.

\begin{figure}[h]
	\centering
	\begin{minipage}{.33\textwidth}
		\centering
		\subfloat[]{\includegraphics[width=5.5 cm,height=4.8 cm]{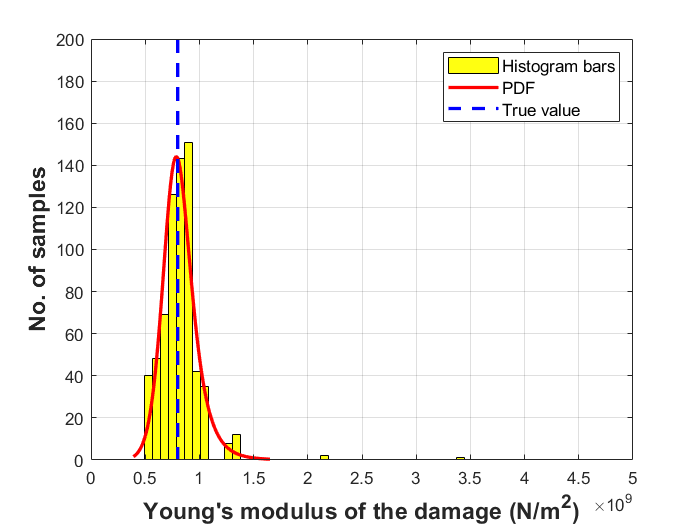}}
		\label{fig: StiffnessHist2}
	\end{minipage}%
	\begin{minipage}{.33\textwidth}
		\centering
		\subfloat[]{\includegraphics[width=5.5 cm,height=4.8 cm]{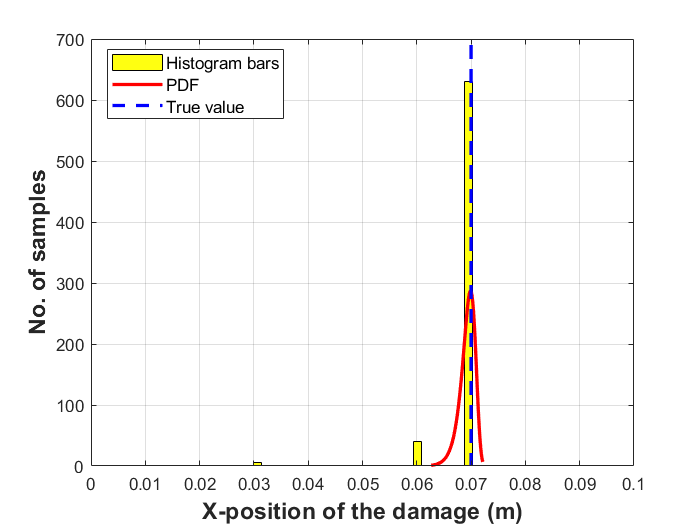}}
		\label{fig: XposHist2}
	\end{minipage}
	\begin{minipage}{.33\textwidth}
		\centering
		\subfloat[]{\includegraphics[width=5.5 cm,height=4.8 cm]{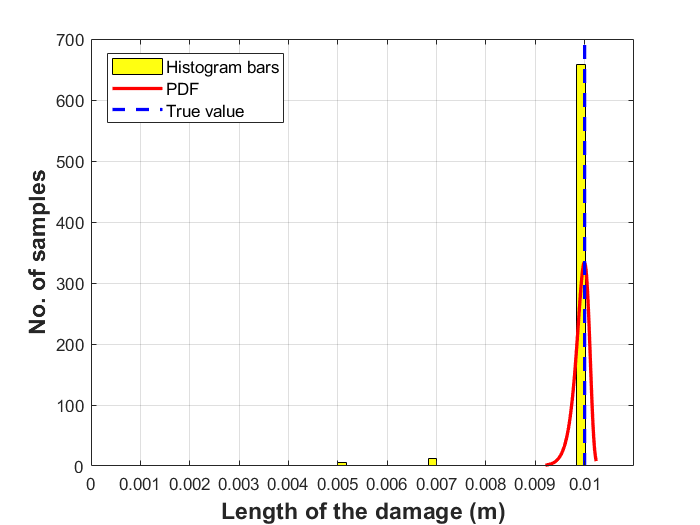}}
		\label{fig: LengthHist2}
	\end{minipage}
	\caption{Histogram of the samples drawn from the posterior distribution of (a) elasticity modulus of the damage (b) position of the damage along the length of the FML structure (c) length of the damage in the structure, for $5\%$ noise level.}
	\label{fig: Histograms2}
\end{figure}

\begin{figure}[h]
	\centering
	\begin{minipage}{.33\textwidth}
		\centering
		\subfloat[]{\includegraphics[width=5.5 cm,height=4.8 cm]{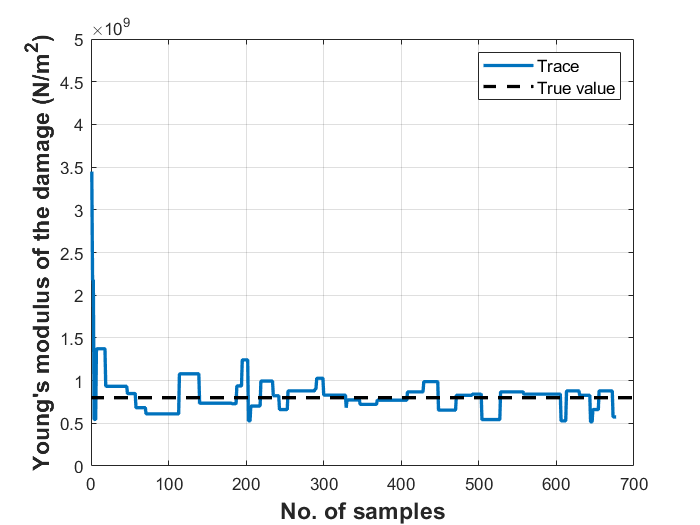}}
		\label{fig: StiffnessTrace2}
	\end{minipage}%
	\begin{minipage}{.33\textwidth}
		\centering
		\subfloat[]{\includegraphics[width=5.5 cm,height=4.8 cm]{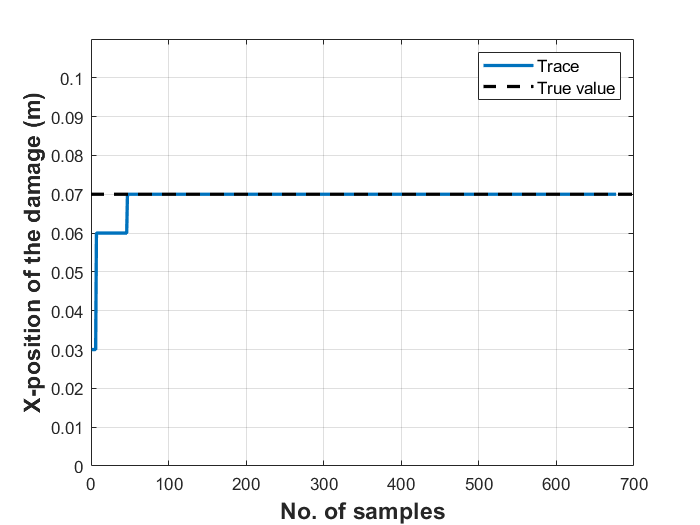}}
		\label{fig: XposTrace2}
	\end{minipage}
	\begin{minipage}{.33\textwidth}
		\centering
		\subfloat[]{\includegraphics[width=5.5 cm,height=4.8 cm]{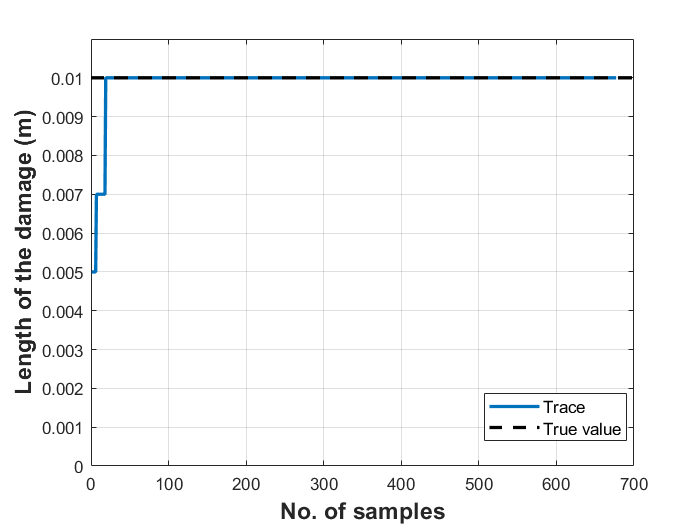}}
		\label{fig: LengthTrace2}
	\end{minipage}
	\caption{Trace of MCMC-MH sampling procedure: (a) elasticity modulus of the damage (b) position of the damage along the length of the FML structure (c) length of the damage in the structure.}
	\label{fig: Trace2}
\end{figure}

Another experiment with a lesser noise amplitude, with $\sigma$ of the error model being $5 \%$ of $u_{d,\max}$, is conducted to study its influence on the inference. The corresponding histograms of the damage parameters and their associated traces in the parametric domain are plotted in Figure~\hspace{-1.5 mm}~\ref{fig: Histograms2} and in Figure~\hspace{-1.5 mm}~\ref{fig: Trace2} respectively. It can be clearly interpreted that processing the data with lower noise amplitude results in reduced uncertainty of the parameters, although considerably less number samples are drawn. This can be clearly seen in the case of Young's modulus of the damage (see Figure \hspace{-1.5 mm}~\ref{fig: Histograms2}(a)). 

\begin{figure}[h]
	\centering
	\begin{minipage}{.33\textwidth}
		\centering
		\subfloat[]{\includegraphics[width=5.5 cm,height=4.8 cm]{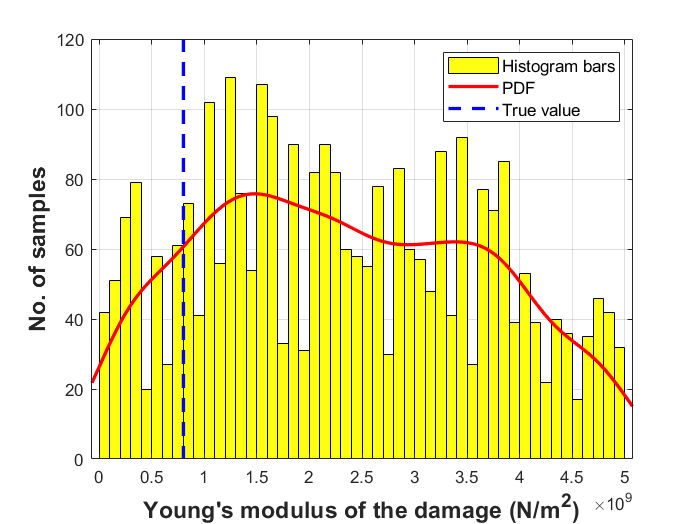}}
		\label{fig: StiffnessHist2_2}
	\end{minipage}%
	\begin{minipage}{.33\textwidth}
		\centering
		\subfloat[]{\includegraphics[width=5.5 cm,height=4.8 cm]{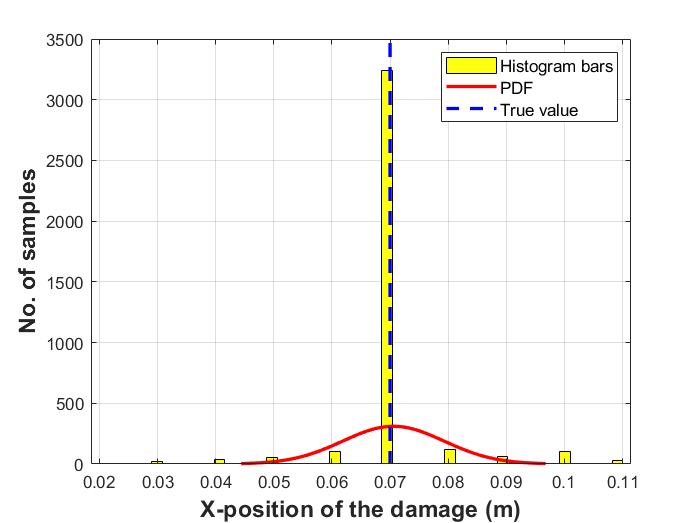}}
		\label{fig: XposHist2_2}
	\end{minipage}
	\begin{minipage}{.33\textwidth}
		\centering
		\subfloat[]{\includegraphics[width=5.5 cm,height=4.8 cm]{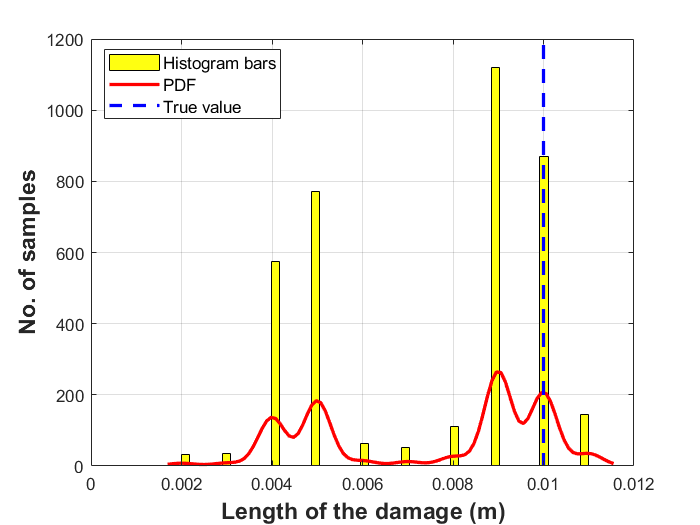}}
		\label{fig: LengthHist2_2}
	\end{minipage}
	\caption{Histogram of the samples drawn from the posterior distribution of (a) elasticity modulus of the damage (b) position of the damage along the length of the FML structure (c) length of the damage in the structure, for $15\%$ noise level.}
	\label{fig: Histograms15_CMAME}
\end{figure}

\begin{figure}[h]
	\centering
	\begin{minipage}{.33\textwidth}
		\centering
		\subfloat[]{\includegraphics[width=5.5 cm,height=4.8 cm]{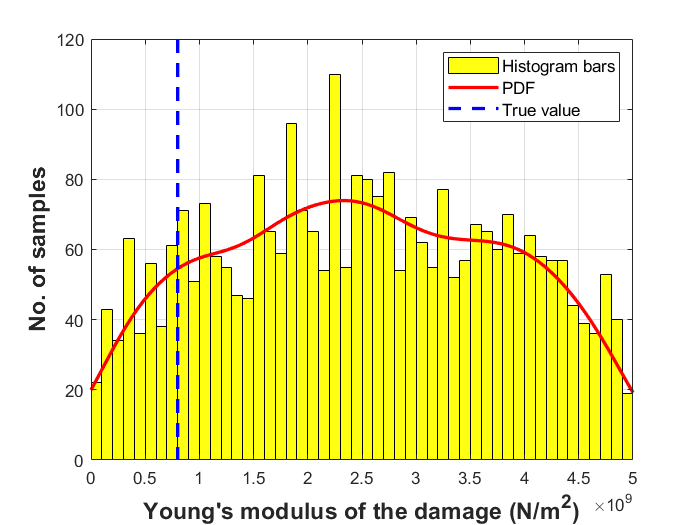}}
		\label{fig: StiffnessHist2_3}
	\end{minipage}%
	\begin{minipage}{.33\textwidth}
		\centering
		\subfloat[]{\includegraphics[width=5.5 cm,height=4.8 cm]{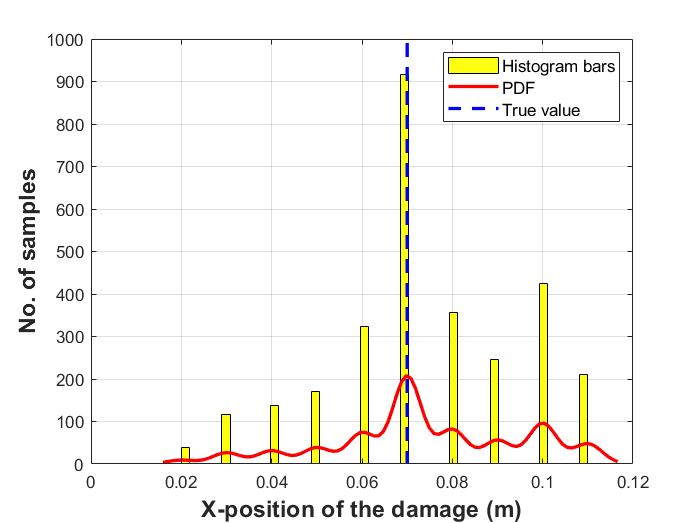}}
		\label{fig: XposHist2_3}
	\end{minipage}
	\begin{minipage}{.33\textwidth}
		\centering
		\subfloat[]{\includegraphics[width=5.5 cm,height=4.8 cm]{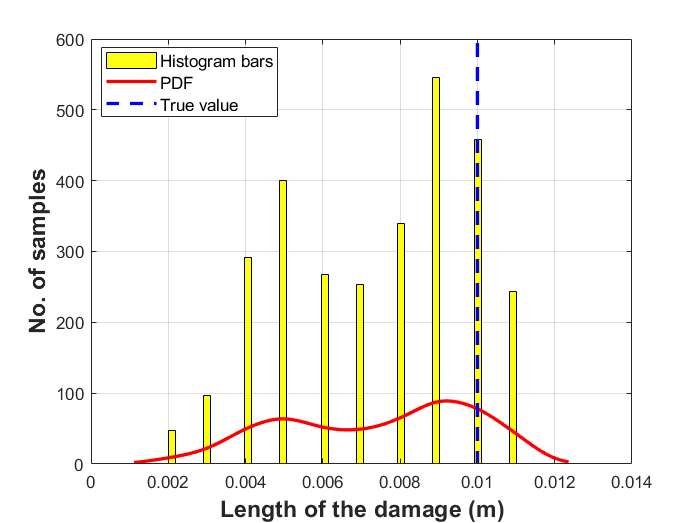}}
		\label{fig: LengthHist2_3}
	\end{minipage}
	\caption{Histogram of the samples drawn from the posterior distribution of (a) elasticity modulus of the damage (b) position of the damage along the length of the FML structure (c) length of the damage in the structure, for $20\%$ noise level.}
	\label{fig: Histograms20_CMAME}
\end{figure}

To understand the limitation of the MCMC-MH algorithm regarding the measurement uncertainty, experiments with higher noise levels of $15\%$ and $20\%$ are conducted. The histograms of the inferred damage parameters to these noise levels are shown in Figure \hspace{-1.5 mm}~\ref{fig: Histograms15_CMAME} and Figure \hspace{-1.5 mm}~\ref{fig: Histograms20_CMAME} respectively. It is evident from the figures that the inference of damage parameters, especially Young's modulus, deteriorates with increased measurement uncertainty. Also, the length of the damage that was previously inferred with high accuracy for the $5\%$ and $10\%$ noise levels has exhibited a bimodal distribution indicating increased uncertainty.

\subsection{Ensemble Kalman filter}
\begin{figure}[h]
	\centering
	\includegraphics[width=15 cm]{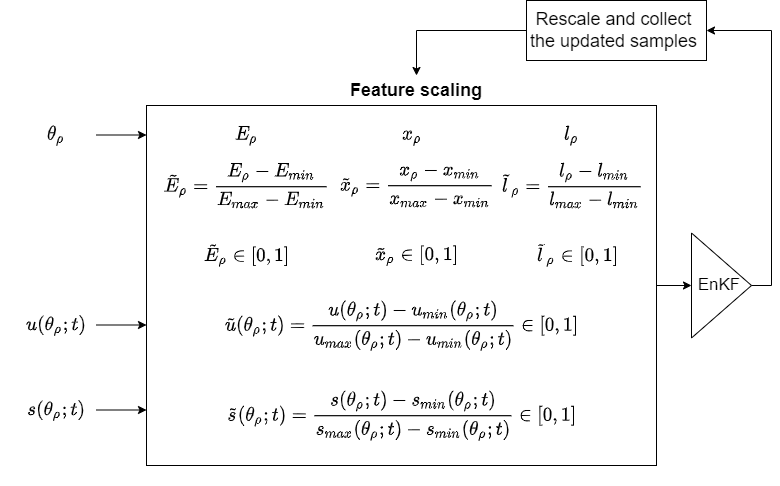}
	\caption{Feature scaling of parameters, model output and observation prior to EnKF}
	\label{fig: FeatureScale}
\end{figure}

As the range of the parameters and the output of the underlying model is far-flung, the computations involved in the prediction-analysis stages of EnKF are non-trivial. Therefore, a data preprocessing technique, feature scaling, is required for the correct computation of sample means, sample covariances, cross-covariances and their updates. In this application, the parameters of the model, system output and the observation of the underlying system are separately normalized between 0 and 1. The normalized values are then used for evaluations in the prediction-analysis steps of the EnKF method. Subsequently, the updated values which lie between 0 and 1 are re-scaled to their original range before starting the next iteration of the EnKF update. The schematic representation of feature scaling of the interested quantities is shown in Figure \hspace{-1.5 mm}~\ref{fig: FeatureScale}. The procedure is carried out using a six particle ensemble i.e. $N_e = 6$. Due to the high frequency of Lamb waves, the resulting solution of the system is a complex wave pattern. This required the filtering technique to update the parameter ensembles much more frequently, at every alternate observation over time ($K=2$ as per Figure 5). 

\begin{figure}[h]
	\centering
	\begin{minipage}{.33\textwidth}
		\centering
		\subfloat[]{\includegraphics[width=5.5 cm,height=4.8 cm]{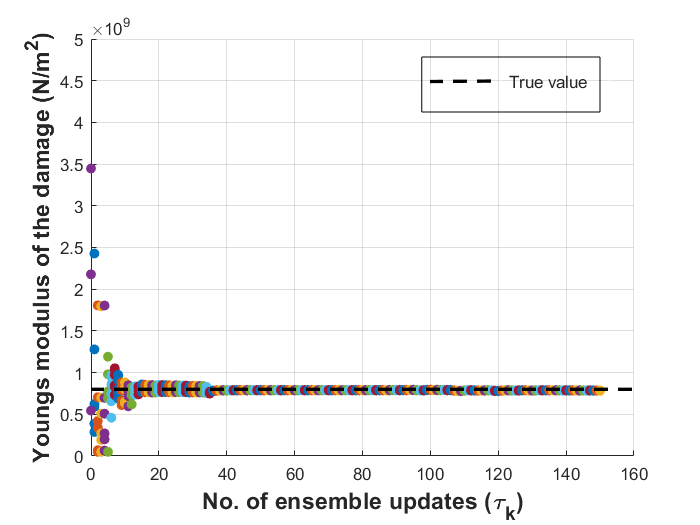}}
		\label{fig:StiffnessHist}
	\end{minipage}%
	\begin{minipage}{.33\textwidth}
		\centering
		\subfloat[]{\includegraphics[width=5.5 cm,height=4.8 cm]{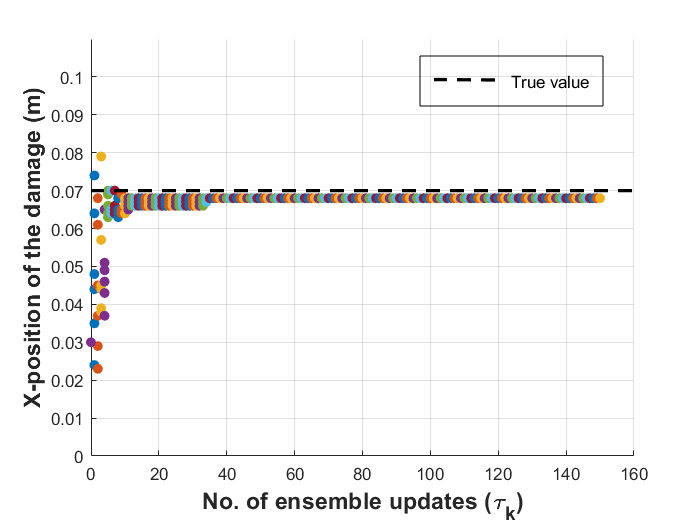}}
		\label{fig:XposHist}
	\end{minipage}
	\begin{minipage}{.33\textwidth}
		\centering
		\subfloat[]{\includegraphics[width=5.5 cm,height=4.8 cm]{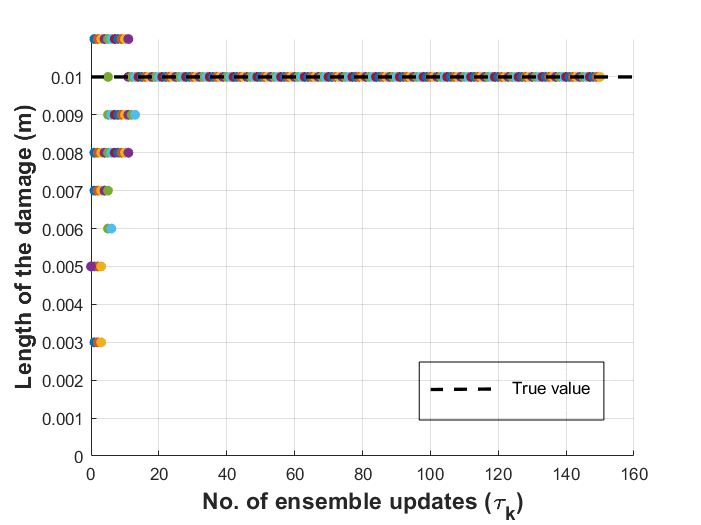}}
		\label{fig:LengthHist}
	\end{minipage}
	\caption{Convergence of parameter ensemble using EnKF: (a) elasticity modulus of the damage (b) position of the damage along the length of the FML structure (c) length of the damage in the structure, for $N_e = 6$ in the ensemble.}
	\label{fig: EnKF_Conv1}
\end{figure}

By using the sensor measurement signal with $\sigma = 10\hspace{1 mm}\%$ of $u_{d,\max}$ and the forward ROM, the EnKF is applied according to Algorithm 2. The particles of the initial ensemble are randomly selected from $\mathcal{D}$. Figure\hspace{-1.5 mm}~\ref{fig: EnKF_Conv1} illustrates the convergence of the ensemble of damage parameters to their corresponding true values as new information flows into the EnKF over time. It is supposed for all the experiments that the excitation signal characteristics are the same and the damage occurs only to the right-hand side of the sensor as shown in Figure \hspace{-1.5 mm}~\ref{fig: ModelSetup}. As the position of the embedded sensor is fixed, the information until the end of the excitation wave package is common for all the parameter configurations. In order to avoid unnecessary computational effort by processing the ambiguous information, observations only after the incident excitation wave package are considered. 

\begin{figure}[h]
	\centering
	\begin{minipage}{.33\textwidth}
		\centering
		\subfloat[]{\includegraphics[width=5.5 cm,height=4.8 cm]{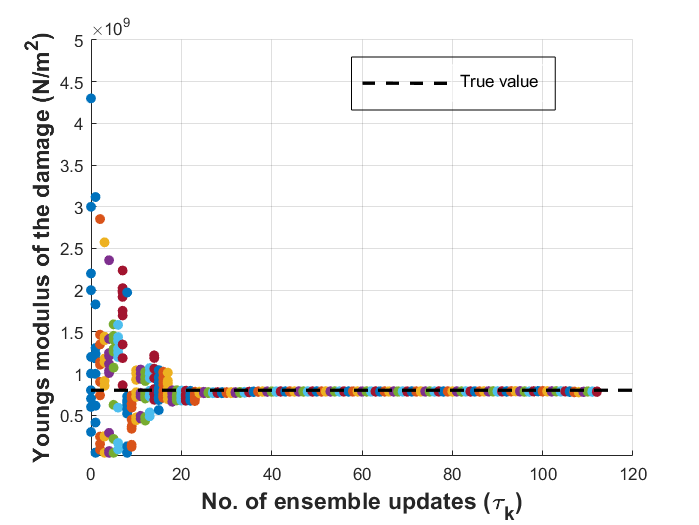}}
	\end{minipage}%
	\begin{minipage}{.33\textwidth}
		\centering
		\subfloat[]{\includegraphics[width=5.5 cm,height=4.8 cm]{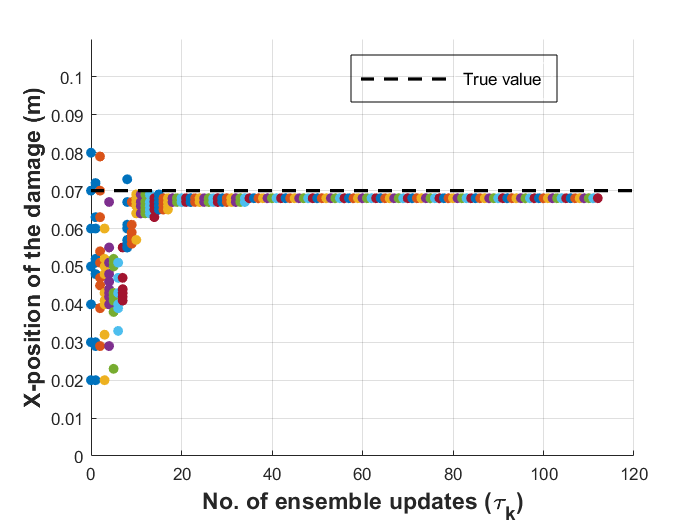}}
	\end{minipage}
	\begin{minipage}{.33\textwidth}
		\centering
		\subfloat[]{\includegraphics[width=5.5 cm,height=4.8 cm]{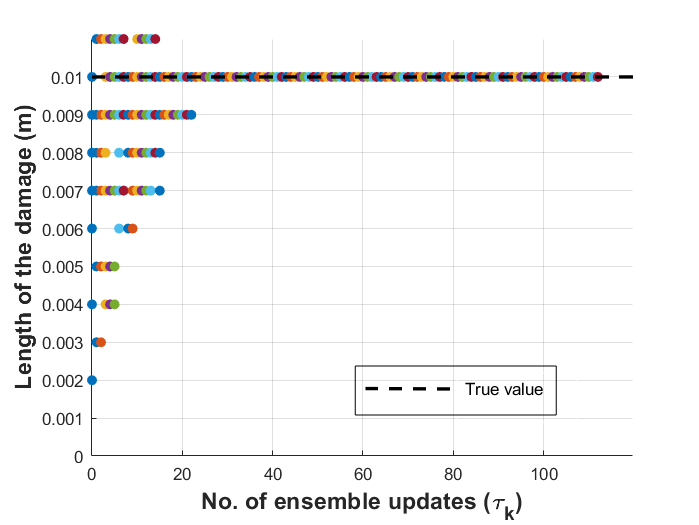}}
	\end{minipage}
	\caption{Convergence of parameter ensemble using EnKF: (a) elasticity modulus of the damage (b) position of the damage along the length of the FML structure (c) length of the damage in the structure for $N_e = 10$ in the ensemble.}
	\label{fig: EnKF_Conv2}
\end{figure}

It is also delved further to probe the effect of an increase in the number of particles in the ensemble, from $N_e = 6$ to $N_e = 10$, on the accuracy and efficiency of parameter estimation by the EnKF method. Figure \hspace{-1.5 mm}~\ref{fig: EnKF_Conv2} demonstrates the convergence of the ensembles to their true values over a period of time. It can be observed from Figure \hspace{-1.5 mm}~\ref{fig: EnKF_Conv1} and Figure \hspace{-1.5 mm}~\ref{fig: EnKF_Conv2} that after 35 and 21 updates of the ensembles respectively, there is no considerable change in the values of the particles and also, they closely converged to the true values of the damage parameters. This sums up to approximately 210 solves of the underlying system in both cases. Therefore, it can be stated in this case, that increasing the number of particles in the ensemble did not improve the efficiency of estimation. 

\begin{figure}[h]
	\centering
	\begin{minipage}{.33\textwidth}
		\centering
		\subfloat[]{\includegraphics[width=5.5 cm,height=4.8 cm]{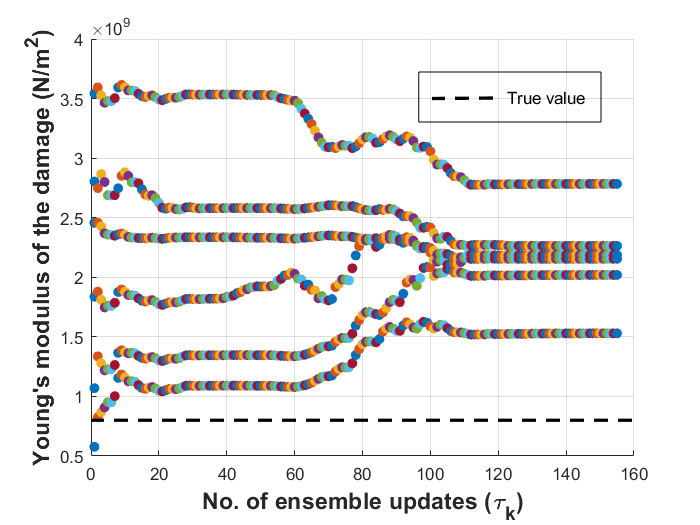}}
	\end{minipage}%
	\begin{minipage}{.33\textwidth}
		\centering
		\subfloat[]{\includegraphics[width=5.5 cm,height=4.8 cm]{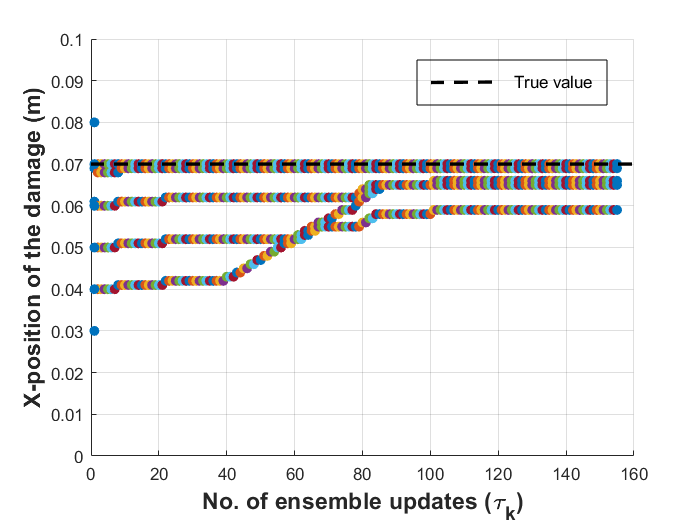}}
	\end{minipage}
	\begin{minipage}{.33\textwidth}
		\centering
		\subfloat[]{\includegraphics[width=5.5 cm,height=4.8 cm]{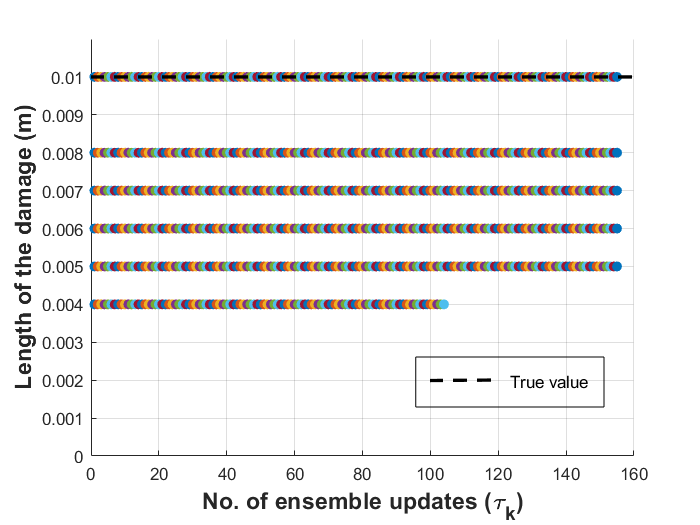}}
	\end{minipage}
	\caption{Convergence of parameter ensemble using EnKF: (a) elasticity modulus of the damage (b) position of the damage along the length of the FML structure (c) length of the damage in the structure for $N_e = 10$ in the ensemble.}
	\label{fig: EnKF_Conv2_2}
\end{figure}

As the noise level is increased to $15\%$ of $u_{d,max}$, the EnKF failed to identify Young's modulus and length of the damage while the parameter ensemble looks converging toward the true value of the damage position (see Figure \hspace{-1.5 mm}~\ref{fig: EnKF_Conv2_2}). This already explains the limitation of this sequential data assimilation algorithm.

\section{Discussions}
\label{sec:discussion}
The above in-silico experiments show that both MCMC-MH and EnKF techniques are capable enough to closely estimate the parameters of the damage in the FML structure. However, for $\sigma = 10\%$ of $u_{d,max}$, the MCMC-MH takes a relatively longer time to converge to the true value, especially for the parameter, Young's modulus of the damage. The time consumed as well as the accuracy corresponding to both methods in identifying the damage, $\boldsymbol{\theta} = \{0.8 \textup{ GPa}, 70 \textup{ mm}, 10 \textup{ mm}\}$, in the FML are encapsulated in Table \hspace{-1 mm}~\ref{table:CompEff} and Table \hspace{-1 mm}~\ref{table:CompAccuracy} respectively. 

\begin{table}[H]
	\caption{Computational effort for damage identification} 
	\centering 
	\begin{tabular}{c c c c} 
		\hline\hline 
		\textbf{Method} & \textbf{No. of forward solves} & \textbf{Computational time} \\ [0.5ex] 
		\hline 
		MCMC-MH & 600 & 5197.2 s\\ 
		EnKF, $N_e = 6$ & 210 & 1789.2 s \\ [1ex] 
		\hline 
	\end{tabular}
	\label{table:CompEff} 
\end{table}

\begin{table}[H]
	\caption{Accuracy of damage identification} 
	\centering 
	\begin{tabular}{c c c c} 
		\hline\hline 
		\textbf{Method} & \textbf{Young's modulus (Pa)} & \textbf{X-position (mm)} & \textbf{Length (mm)}\\ [0.5ex] 
		\hline 
		MCMC-MH & 837934000 & 70 & 9.7\\ 
		EnKF, $N_e = 6$ & 786310000 & 68 & 10 \\ 
		True value & 800000000 & 70 & 10 \\[1ex] 
		\hline 
	\end{tabular}
	\label{table:CompAccuracy} 
\end{table}

\subsection{MCMC-MH algorithm}

\begin{figure}[H]
	\centering
	\begin{minipage}{.33\textwidth}
		\centering
		\subfloat[]{\includegraphics[width=5.5 cm,height=4.8 cm]{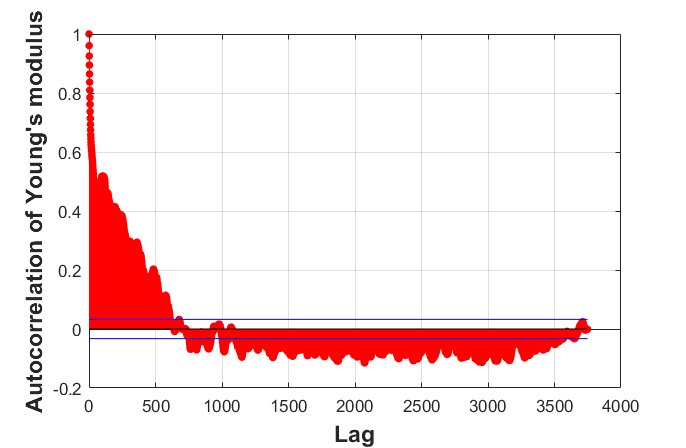}}
		\label{fig: Auto_E}
	\end{minipage}%
	\begin{minipage}{.33\textwidth}
		\centering
		\subfloat[]{\includegraphics[width=5.5 cm,height=4.8 cm]{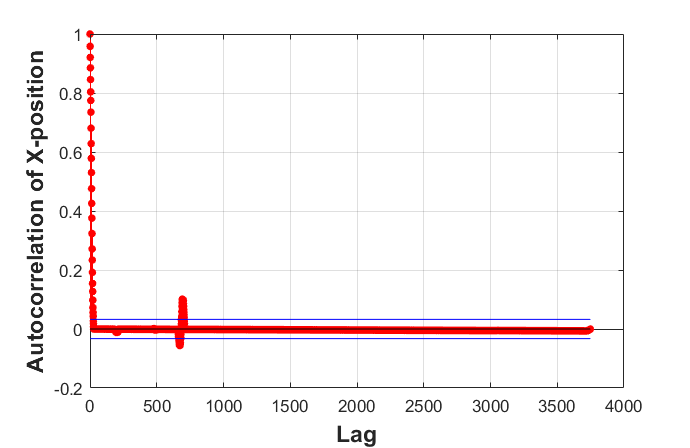}}
		\label{fig: Auto_X}
	\end{minipage}
	\begin{minipage}{.33\textwidth}
		\centering
		\subfloat[]{\includegraphics[width=5.5 cm,height=4.8 cm]{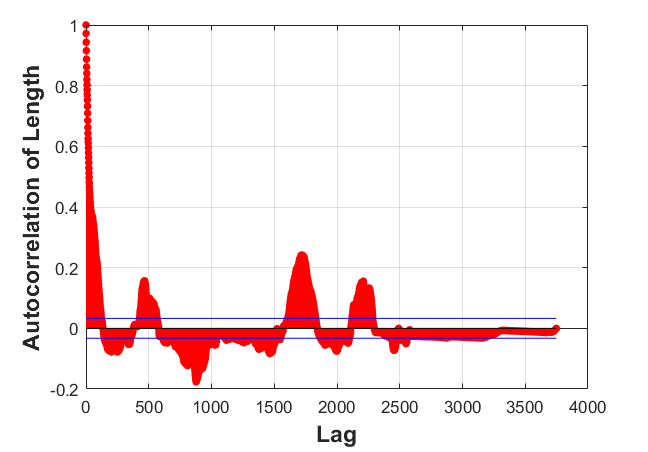}}
		\label{fig: Auto_L}
	\end{minipage}
	\caption{Sample autocorrelation function for: (a) Young's modulus (b) position along the length of the FML (c) length of the damage in FML.}
	\label{fig: Autocorrelation}
\end{figure}

As informed in Section 4.2, the MCMC algorithms draw a sample that depends on the previous sample. This induces a correlation between the samples drawn from the posterior distribution. However, as the number of samples drawn becomes sufficiently larger, they become almost independent. If the effective MCMC sample size is too small, then the corresponding distribution of these samples usually lies far from the target distribution. This is potentially diagnosed by plotting the correlogram of the parameters. Figure \hspace{-1.5 mm}~\ref{fig: Autocorrelation} illustrates the decay of autocorrelation with increasing lag and that it is bounded within the $95\%$ confidence band of the drawn number of samples shown in solid blue lines. This substantiates the number of samples drawn and the accuracy of the empirical distributions shown in Figure \hspace{-1.5 mm}~\ref{fig: Histograms1}.

\begin{figure}[H]
	\centering
	\includegraphics[width=9 cm]{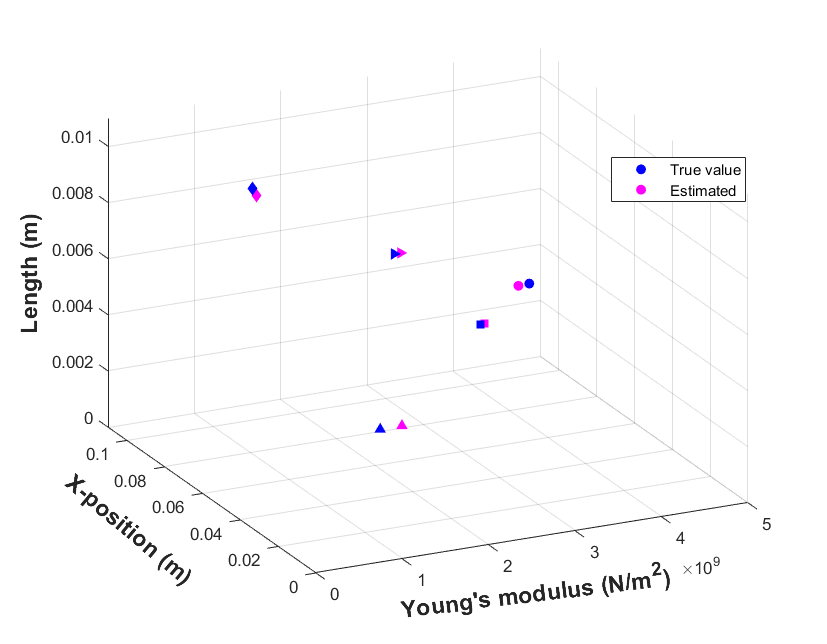}
	\caption{Accuracy of MCMC-MH algorithm for different damage cases}
	\label{fig: Robustness_MCMC}
\end{figure}

In order to corroborate the robustness of the MCMC-MH algorithm for this application, four more experiments for different damage parameters are conducted. It is found that the MCMC-MH procedure could infer the damage with high accuracy. After discarding the burn-in samples, the mean of the parameters are plotted in a three-dimensional scatter plot along with their corresponding true values (see Figure \hspace{-1.5 mm}~\ref{fig: Robustness_MCMC}).  Each marker symbol represents an experiment wherein the true values are plotted in blue color while the mean of the samples excluding those in the burn-in period is in magenta.

\subsection{Ensemble Kalman filter}
\begin{figure}[H]
	\centering
	\begin{minipage}{.33\textwidth}
		\centering
		\subfloat[]{\includegraphics[width=5.5 cm,height=4.8 cm]{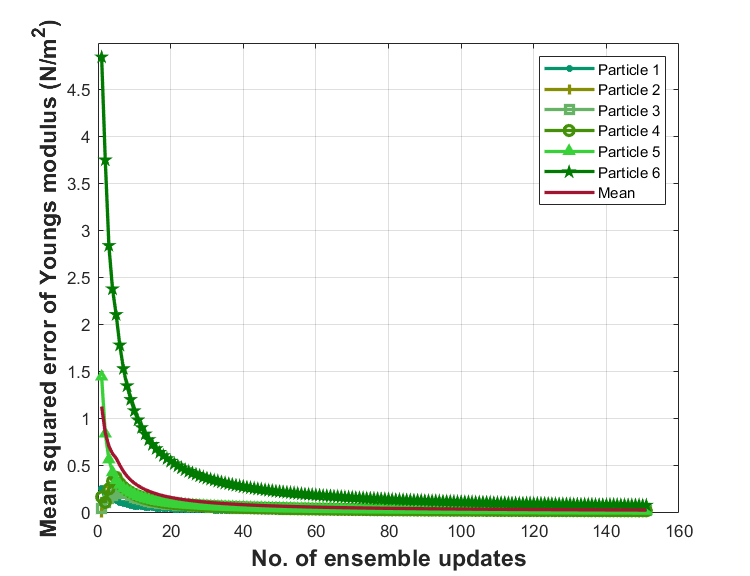}}
		\label{fig: MAE}
	\end{minipage}%
	\begin{minipage}{.33\textwidth}
		\centering
		\subfloat[]{\includegraphics[width=5.5 cm,height=4.8 cm]{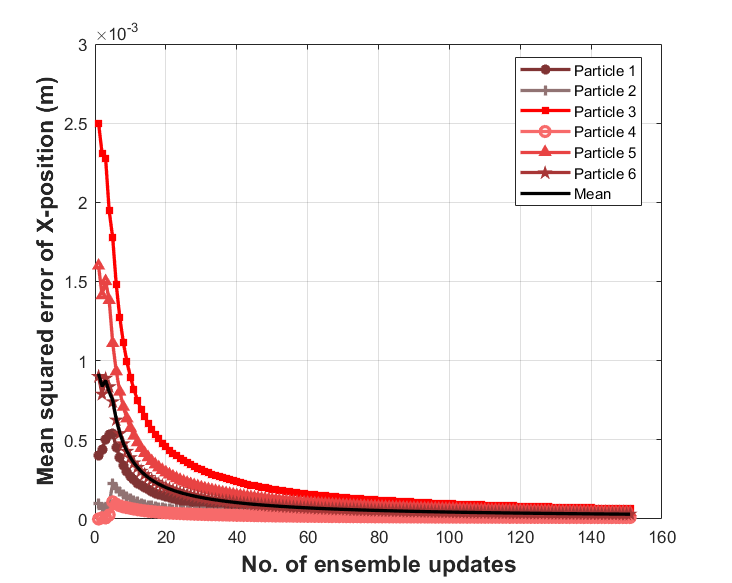}}
		\label{fig: MSEX}
	\end{minipage}
	\begin{minipage}{.33\textwidth}
		\centering
		\subfloat[]{\includegraphics[width=5.5 cm,height=4.8 cm]{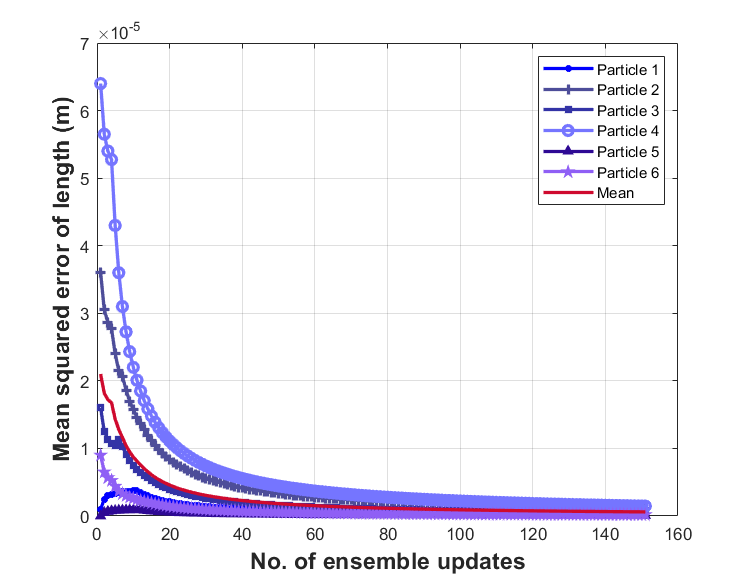}}
		\label{fig: MSEL}
	\end{minipage}
	\caption{Mean squared error of particles in the ensemble: (a) Young's modulus (b) position along the length of the FML (c) length of the damage in FML.}
	\label{fig: ErrConv}
\end{figure}

The evolution of the mean squared error of the ensemble with 6 particles is plotted in Figure\hspace{-1.5 mm}~\ref{fig: ErrConv}, to study its convergence. It is clearly evident that as the new information flows into the EnKF algorithm, the error corresponding to the particles approaches zero indicating the convergence to the true values of the parameter.  

\begin{figure}[H]
	\centering
	\includegraphics[width=9 cm]{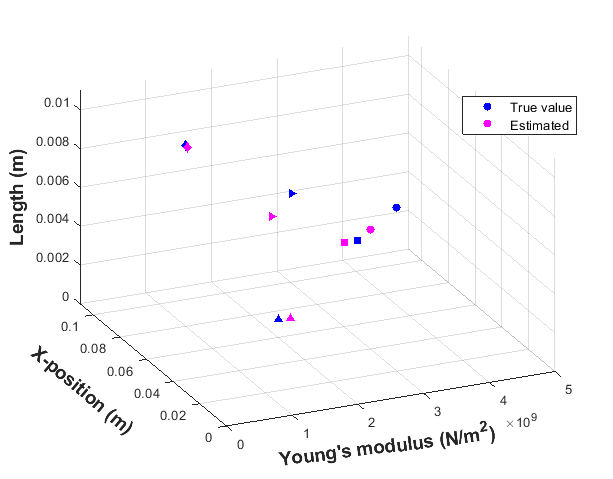}
	\caption{Accuracy of EnKF technique corresponding to different damage cases}
	\label{fig: Robustness}
\end{figure}

Several experiments for different damage cases are conducted and the robustness of the employed methods concerning the damage identification in FML is analyzed. Figure\hspace{-1.5 mm}~\ref{fig: Robustness} indicates how well the EnKF technique is able to characterize damages of different parameters. Each marker symbol represents an experiment wherein the true values are plotted in blue color while the mean of the latest 100 ensemble particles is in magenta.

\section{Conclusion}
\label{sec:conlusion}
Damage identification at an earlier stage is crucial to prevent catastrophic failures in layered materials that are employed in aerospace structures and wind turbine rotor blades. This work investigated the application of Markov chain Monte Carlo (MCMC) based Metropolis-Hastings (MH) algorithm and Ensemble Kalman filtering (EnKF) technique in a stochastic Bayesian paradigm for damage identification within a carbon fiber reinforced epoxy-steel laminate. Since solving the high-fidelity model of the underlying system is cost-intensive, the inference procedure rather utilizes the reduced-order model (ROM) built upon the proper orthogonal decomposition method. The accuracy and efficiency of the ROM can be meticulously seen in \cite{Bellam-Muralidhar-NK} and is also reflected in the results of parameter inference presented in this paper. 

The numerical experiments demonstrate that both the MCMC-MH algorithm and EnKF technique accurately estimate the damage parameters for low noise levels, however, the accuracy exhibited by the MCMC-MH method is relatively higher than that of the EnKF approach (compare Figure~\ref{fig: Robustness_MCMC} and Figure~\ref{fig: Robustness}). Due to the provision of avoiding redundant information, in this particular application, the EnKF method outperforms the MCMC-MH algorithm in terms of computational cost. The EnKF took approximately one-third of the time expended by the MCMC-MH method to infer the parameters from the noisy measurement. As the considered problem is much simpler than the usual complex damage that is likely to appear in the layered laminates, these straightforward methods were able to accurately identify the damage for the considered test case. However, in case of multiple sophisticated damages, advanced algorithms \cite{He-S, Green, Chiachio, Chiachio-M, Feng-Z, Cheung-SH} should be explored to characterize them.  

In the future, this research work will be extended for profound investigations to account for non-linearity that will be introduced due to material behavior or contact between the delamination interfaces. The global ROMs suffer from the disadvantage that the reduced dimension grows as the size or dimension of the parametric space increases which ultimately adds a huge computational burden. That being the case, we will also seek to produce several ROMs that are associated with specific localized parametric domains \cite{Son-NT, Choi-Y-LROM, Zimmermann-RMIR} which could increase the accuracy and efficiency. Furthermore, instead of the synthetically generated measurements, real-world sensor data from an in-situ structural health monitoring (SHM) system will be utilized to identify the damage. As the number of sensors integrated into the SHM system is minimal, a judicious positioning of these sensors in the structure is vital. Therefore, this research will pursue optimizing the sensor positions, in the case of multiple sensors, to maximize the value of information gain from the system for a more accurate identification of the damage. The value of information gain will be the difference between the prior and posterior PDF of the damage parameters. The existing Bayesian framework will be used to compute the information gain and an optimizer will be wrapped on this setup to find the optimal sensor position \cite{Konakli, Papadimitriou, Cantero, Cantero2020}. Finally, model uncertainty has not been taken into account in this work, and could be a topic for further investigation.

\section*{Acknowledgements}     

The authors Bellam Muralidhar, Rauter, Mikhaylenko, Lammering and Lorenz acknowledge the financial support of this research work within the Research Unit 3022 “Ultrasonic Monitoring of Fiber Metal Laminates Using Integrated Sensors” by the German Research Foundation (Deutsche Forschungsgemeinschaft (DFG)) under grant numbers LO1436/12-1, RA 3433/1-1 and 418311604.

\section*{References}

\bibliography{Bib_Ref}
\bibliographystyle{plain}

\end{document}